\documentstyle[prb,preprint,aps,epsf]{revtex}

\begin{document}

\newcommand{\bc}{\begin{center}}
\newcommand{\ec}{\end{center}}
\newcommand{\ba}{\begin{array}}
\newcommand{\ea}{\end{array}}
\newcommand{\bt}{\begin{tabular}}
\newcommand{\et}{\end{tabular}}
\newcommand{\bi}{\begin{itemize}}
\newcommand{\ei}{\end{itemize}}
\newcommand{\be}{\begin{equation}}
\newcommand{\ee}{\end{equation}}
\newcommand{\beqn}{\begin{eqnarray}}
\newcommand{\eeqn}{\end{eqnarray}}
\newcommand{\binomial}[2]{\biggl(\ba{c} #1 \\ #2 \ea\biggr)}
\newcommand{\cala}{{\cal  A}}
\newcommand{\calb}{{\cal  B}}
\newcommand{\ds}{\displaystyle}

\title{The One-Dimensional ANNNI model in a Transverse Field:\\
Analytic and numerical study of Effective Hamiltonians}

\author{Heiko Rieger$^{1,2}$ and Genadi Uimin$^{1,3}$}

\address{
$^1$ Institut f\"ur Theoretische Physik, 
     Universit\"at zu K\"oln, 50937 K\"oln, Germany\\
$^2$ HLRZ, Forschungszentrum J\"ulich, 52425 J\"ulich, Germany\\
$^3$ Landau Institute for Theoretical Physics, 
     Chernogolovka, Moscow District, Russia
}
 
\date{\today} 

\maketitle

\begin{abstract} 
We consider a spin-$\frac{1}{2}$ chain with competing nearest and
next-nearest neighbor interactions within a transverse magnetic field,
which is known to be an equiavelent to the ANNNI model. When studing 
thermodynamics of the 2D ANNNI model Villain and
Bak arrived to a free fermion approximation that neglects heavy
excitations from the ferromagnetic ground state, which is an
appropriate description close to the paramagnetic-ferromagnetic
transition. In the vicinity of the floating-phase/anti-phase transition
another sort of quasiparticles, but free fermions too, appears to be 
convenient. Although free fermions are a suitable tool for investigation
of the phase diagram and the critical properties, they are defined on the
fictitious lattice which makes the analysis non-rigorous.
Here we deal with a proper fermion scheme which is especially effective 
%devised to describe the floating-phase/anti-phase transition. 
for performing exact diagonalization calculations for cyclic chains.
Systems up to size $L=32$ has been analysed and the
predictions of the effective fermion Hamiltonian has been confirmed. 
Various predictions for the infinite system and the critical properties are
derived.
\end{abstract}


\section{Introduction}

Frustrated Ising models within a transverse field have been
investigated for a long time \cite{Chakrabarti}. The transverse field
plays the role of a tunable parameter by which one can induce a so
called ``quantum phase transition'' at zero temperature that is driven
by quantum fluctuation alone \cite{QPT} (as opposed to a conventional,
thermally driven phase transition). Frustration can be introduced via
disorder, as for example in the case of quantum Ising spin glasses
that have gained much interest quite recently \cite{QISG}. However, it
can also be produced in a regular fashion as e.g.\ in the anisotropic
next nearest neighbor Ising model (ANNNI) \cite{Selke}, where competing
nearest and next-nearest neighbor interactions are the origin of the
richness of the phase diagram. This is the model that we intend to
re-investigate in this paper. To be concrete, we consider the
Hamiltonian, ${\cal H}$, of a 1$d$ spin-$\frac{1}{2}$ chain in a
transverse magnetic field, which consists of the two parts: the classical
1$d$ ANNNI model plus quantum fluctuations imposed by transverse
field:
\be
{\cal H}=H_{{\rm cl}}+H_{{\rm qu}},
\label{ham}
\ee
\be
\label{mast_eq}
H_{{\rm cl}}=-{\cal J}\sum_i\sigma_i^z\sigma_{i+1}^z+
\kappa {\cal J}\sum_i\sigma_i^z\sigma_{i+2}^z
\quad{\rm and}\quad
H_{{\rm qu}}=-\Gamma\sum_i\sigma_i^x
\ee
$\Gamma=0$ corresponds to a classical ANNNI model 
in which ${\cal J}$ is supposed to be
positive. $\sigma^z$ and $\sigma^x$ are the Pauli matrices:
\beqn
\sigma^z=\left(
\begin{array}{cc}
1 & 0\\
0 & {\rm -}1
\end{array}\right),\;\;\;\sigma^x=\left(
\begin{array}{cc}
0 & \;1\\
1 & \;0
\end{array}\right)
\eeqn
The most interesting region of this model is the region around
$\kappa=1/2$. The classical part $H_{\rm cl}$ has an infinite ground
state degeneracy at this point, separating the ferromagnetic (FM) region
($\kappa<1/2$), where the ground-state is given by ``all-spins-up'' or
``all-spins-down'', from the anti-phase region ($\kappa>1/2$), where
the ground-state has a period 4 with two down-spins following two
up-spins. After \cite{fs} $<\!2\!>$ is a traditional notation for this 
anti-phase. Switching on the quantum fluctuations via a nonvanishing
transverse field induces the presence of different phases: fixing
$\Gamma\ll J$ and increasing $\kappa$ in the vicinity of 1/2, one can 
distinguish 
\bi
\item
FM ($\langle \sigma_i^z\rangle\neq 0$); 
\item
PM, {\it i.e.}, paramagnetic phase (exponentially
decaying spatial spin-spin correlations, no long-range order); 
\item
FP, {\it i.e.}, floating phase 
(algebraically decaying spatial spin-spin correlations, accompanied with 
a modulation continuously changed with $\kappa$, no long-range order);
\item
$<\!2\!>$ ($\langle \sigma_i^z \sigma_{i+2}^z\rangle<0$, 
$\langle \sigma_i^z \sigma_{i+4}^z\rangle>0$). 
\ei
The FM--PM transition is well understood: It is within the same
universality class as the two-dimensional Ising model or 1d
unfrustrated Ising model in a transverse field (see, e.g.,
\cite{matt,schultz}).

Villain and Bak \cite{vill} in their seminal work on the
two-dimensional ANNNI model argued that the $<\!2\!>$--FP
transition is expected to be of the Prokovsky-Talapov type \cite{PT}
and that the FP--PM transition is expected
to be of the Kosterlitz-Thouless type \cite{KT}. Since this work is based
on various plausible but not rigorously proven assumptions 
many attempts have been
done to check these predictions with Monte Carlo simulations or exact
diagonalization studies (for a review see \cite{Selke}). However, for
principle reason that we also try to clarify in this paper, such an
endevour turns out to be very difficult and fails to provide the
theory either with a conclusive support or with a clear falsification.

A principal question of this work is how to catch the essential physics 
in the vicinity of the FP--$<\!2\!>$ and FP--PM transitions. For doing this 
we present a regular expansion of Hamiltonian (\ref{ham}) in powers 
$\Gamma/{\cal J}$, which allows to perform exact numerical diagonalization
for longer chains than it is possible to achieve by a straightforward 
procedure. 

The organization of the paper is as follows: In Section II we
recapitulate the free fermion picture of Villain and Bak \cite{vill}
before we present the above mentioned effective
Hamiltonian. Then in Section III we compare the analytical predictions of this 
theory for {\it finite} systems with the results of exact diagonalization
studies and proceed to extract the
desired information about the critical behavior.
% of the {\it infinite system} from our effective Hamiltonian. 
Section IV summarizes our results.

\section{Free fermion picture}

Near $\kappa\!=\!1/2$ the elementary excitations can be distinguished
as light and heavy. The main idea is to map the initial Hamiltonian
onto the states where the heavy excitations are excluded from. 

We start with a conditional FM vacuum state, for example,
$..+\,+\,+\,+\,+..$ , then the excitation with an isolated ($-$) spin,
e.g., $..+\,+\,-\,+\,+..$ costs the energy $4{\cal J}(1-\kappa)$ which
is not small as $\kappa\rightarrow 1/2$. However, the excitation with
two or more sequential spins rotated, e.g., $..+\,+\,-\,-\,+\,+..$ costs the
energy $4{\cal J}(1-2\kappa)$ which vanishes with $\kappa -1/2\,$. 
At the next step we can introduce
quasiparticles, which are the domain walls (DW's) defined on the {\it dual}\
lattice. The latter coincides in 1$d$ with the middles of the
links. The energy of a single DW, {\it i.e.}, 
$..+\,+\,+\!\!\stackrel{\rm{dw}}{|}\!\!-\,-\,-..$ , is
determined by the classical part of ${\cal H}$: 
\be
\epsilon=2{\cal J}(1-2\kappa)\;. 
\ee
The DW's which occupy the nearest sites of the dual lattice repel each
other with the cost of energy $V=4{\cal J}\kappa$.  In general any
state now is characterized by positions of the DW's on the dual
lattice
\be
\nonumber
\{\ell_1,\ell_2,\dots,\ell_k\}.
\ee
For convenience we set the coordinates on a dual lattice to be integer
numbers, $1,\dots,L$, while the sites of a real lattice run
half-integer numbers, say, $1/2,\dots,L-1/2.$ The quantum part of
${\cal H}$ plays a role of the kinetic energy of quasiparticles. In
fact, applying $H_{{\rm qu}}$ to the state with a DW located at the
site $\ell$, the spin from either its right or its left is changed by
sign, that means a shift of a DW by one unit.  Applying $H_{{\rm qu}}$
to the site with no DW's surrounding it creates a couple of DW's.
Hermitian conjugation corresponds to annihilation of those DW's.

All the matrix elements described in terms of DW variables on the dual lattice
can be summarized in the following Hamiltonian:
\beqn
\label{red_1}
{\cal H}=H_0+H_1,\quad H_0=
V\sum_jn_j^{(\tau)}n_{j+1}^{(\tau)}\nonumber\\
H_1=\epsilon\sum_j n_j^{(\tau)}-\Gamma\sum_j(\tau_j^+\tau_{j+1}^++\tau_j^-\tau_{j+1}^-+
\tau_j^+\tau_{j+1}^-+\tau_j^-\tau_{j+1}^+)
\eeqn
where $\tau$'s are usual Pauli matrices and $n^{(\tau)}=(1-\tau^z)/2$. 
A conditional vacuum is a state with no DW's ($\tau_j^z=1$ or 
$n_j^{(\tau)}=0$ in 
(\ref{red_1})), creation (annihilation) of a DW is realized by $\tau^-$
($\tau^+$). A standard derivation of Hamiltonian (\ref{red_1}) and 
transformation from spin-operators $\sigma$'s to spin-operators $\tau$'s
is given in Appendix A.

A routine procedure of the Jordan-Wigner transformation
$$
\tau_j^+=c_j\exp\left(\imath \pi\sum_{k=1}^{j-1}n_k\right)
$$
allows to deal with fermionic variables.
Below we shall use the periodic boundary conditions, that means that
the $L\!+\!1$-th site should be identified with the first. 
Hamiltonian (\ref{red_1}) can be rewritten now as
\beqn
{\cal H}=\sum_{j=1}^{L-1}(\epsilon n_j+Vn_jn_{j+1}-\Gamma
(c^{\dag}_jc^{\dag}_{j+1}+c_{j+1}c_j+c^{\dag}_jc_{j+1}+c^{\dag}_{j+1}c_j))+
\nonumber \\
\label{red_2}
+Vn_1n_L-\Gamma(c^{\dag}_1c^{\dag}_L+c_Lc_1-c^{\dag}_1c_L-c^{\dag}_Lc_1)
\exp (\imath \pi k)
\eeqn
where $k=\sum_1^Ln_j$, a total number of DW's. 
Evidently, $k$ should be even number on a cyclic chain that results in 
$\exp (\imath \pi k)\equiv 1$. 

Assuming $V\!\gg \!(\epsilon,\Gamma)$ one can obtain the effective Hamiltonian,
${\cal H}_{{\rm eff}}^{(1)}$, which reflects low energy properties of 
${\cal H}$
with the energy scale of order  $(\epsilon,\Gamma)$. Note, that the $V$-term
makes two DW's energetically unfavorable if they occupy the nearest sites.
Simultaneously, terms $\Gamma(\tau^+\tau^++\tau^-\tau^-)$ of 
(\ref{red_1}) (or terms $\Gamma(c^{\dag}c^{\dag}+cc)$ of (\ref{red_2}))
should be excluded from the effective Hamiltonian, which now reads in
$\tau$-variables
\beqn
\label{eff_1t}
{\cal H}_{{\rm eff}}^{(1)}=\sum_{j=1}^{L-1}(\epsilon n_j^{(\tau)}-
\Gamma(\tau_j^+\tau_{j+1}^-+\tau_j^-\tau_{j+1}^+))-
\Gamma(\tau_1^+\tau_L^-+\tau_1^-\tau_L^+)
\eeqn
and in fermionic variables
\beqn
\label{eff_1c}
{\cal H}_{{\rm eff}}^{(1)}=\sum_{j=1}^{L-1}(\epsilon n_j -
\Gamma(c^{\dag}_jc_{j+1}+c^{\dag}_{j+1}c_j))+
\Gamma(c^{\dag}_1c_L+c^{\dag}_Lc_1).
\eeqn
The constraint should be imposed on a possible wave function: 
DW's, or fermions, occupying the nearest sites, 
are forbidden.

Let us try a wave function of the form 
\be
\label{wf1}
\psi=\sum_{m_1}\dots\sum_{m_k}f(m_1,m_2,\dots,m_k)c^{\dag}_{m_1}
c^{\dag}_{m_2}\dots c^{\dag}_{m_k}
\ee
where $k$ should be even.
We may search for the amplitudes $f$'s in a form of Bethe
substitution:
\be
\label{bethe}
f(m_1,m_2,\dots,m_k)=\sum_{\{P\}}\xi_P\exp \imath(q_{P_1}m_1+\dots+q_{P_k}m_k)
\ee
where $\{P\}$ is a permutation of numbers $\{1,2,\dots,k\}$.
Using the results of Appendix B we obtain a general expression for
the eigenfunctions of Hamiltonian (\ref{red_2}) in the $V\rightarrow\infty$
limit
\beqn
\label{wf}
\psi=\sum_{m_1}\dots\sum_{m_k}\left|
\ba{cccc}
e^{\imath q_1m_1} & e^{\imath q_1(m_2-1)} & \cdots & e^{\imath q_1(m_k-k+1)}\\
e^{\imath q_2m_1} & e^{\imath q_2(m_2-1)} & \cdots & e^{\imath q_2(m_k-k+1)}\\
\cdots & \cdots & \cdots & \cdots \\
e^{\imath q_km_1} & e^{\imath q_k(m_2-1)} & \cdots & e^{\imath q_k(m_k-k+1)}
\ea
\right|c^{\dag}_{m_1}
c^{\dag}_{m_2}\dots c^{\dag}_{m_k}|0\rangle.
\eeqn
The wave function (\ref{wf}) can be interpreted as a wave function of 
a tight-binding fermion model on a fictitious lattice. The coordinate of a
fermion in a fictitious lattice coincides with that one in a real lattice
minus the number of fermions situated from its left. We shall also use 
another interpretation which will be convenient in numerical diagonalization.
It is consistent with introducing two kind of ``particles'', ${\cal A}$ 
and ${\cal B}$. ${\cal A}$ is composed of a DW (or fermion) with a nearest
empty  
site from its right attached. ${\cal B}$ represents an empty site which has 
no a nearest DW from its left. The ${\cal A}-{\cal B}$ representation will be
discussed in Section III in detail. 

The ground state energy
of any intermediate state on the phase diagram between the FM and $<\!2\!>$ 
boundaries
\be
\label{gse}
E_{{\rm gs}}=k\epsilon -4\Gamma\sum_{m=1}^{k/2}\cos \frac{(2m-1)\pi}{L-k}
=k\epsilon-2\Gamma\frac{\sin \pi k/(L-k)}{\sin \pi/(L-k)}
\ee
is a function of $k$, the total amount of DW's, which characterizes
modulation of a spin structure.
Eq.(\ref{gse}) describes the ground-state energies of the FM structure 
($k=0, E_{{\rm gs}}=0$) and of the $<\!2\!>$  structure 
($k=L/2, E_{{\rm gs}}= L\epsilon/2$) as well. Both structures, FM and 
$<\!2\!>$, 
can be unstable with respect to formation either DW's ($k=2$) 
or ``holes'' in a regular
DW structure ($k=L/2-2$)\cite{footnote}
%\footnote{For the infinite system it will be enough
%to check $k=1$ and $k=N/2-1$, respectively, but for a periodic system
%we must use even $k$'s.}

For a finite cyclic chain we determine the boundaries from equations:
\bi
\item
FM--FP: \( \epsilon=2\Gamma\cos \frac {\pi}{L-2}
\stackrel{L\rightarrow\infty}{\Longrightarrow}2\Gamma\)
\item
$<\!2\!>$--FP: $\epsilon=-2\Gamma\left(\cos\frac {2\pi}{L+4}+
\cos\frac {6\pi}{L+4}\right)\stackrel{L\rightarrow\infty}{\Longrightarrow}
-4\Gamma$
\ei
For an infinite chain we may introduce $q=\lim_{L\rightarrow\infty}k/L$,
the concentration of DW's, which may be varied from 0 to 1/2. The analogue of 
(\ref{gse}) in the limit $L\rightarrow\infty$
\be
\frac {E_{{\rm gs}}}L=q\epsilon-2(1-q)\Gamma\int_0^{k_F}\frac{dk}{\pi} \cos k
\label{gseinf}
\ee
where $k_F=\pi q/(1-q)$. Differentiating over $q$ in (\ref{gseinf}) shows 
how $q$ changes with $\epsilon/\Gamma\,$ \cite{vill}:
\be
\frac\epsilon\Gamma=-\frac 1\pi\sin k_F +\frac 1{1-q}\cos k_F\;.
\label{qeq}
\ee

The first excited states whose energies are numerically calculated 
in Section III  are non-trivial even in the
framework of the free fermionic approach (cf Eqs.(\ref{e1}) and (\ref{e2})). 
In Appendix C we derive Eq.(\ref{en_exc}) for the
energy $\Delta E(Q)$ as a function of the wavevector $Q$.
Such a fermion-hole excitation leaves the DW number
unchanged. It becomes gapless at certain $Q$'s when $L\rightarrow\infty$.
In this limit the only non-zero contribution to $\Delta E(Q)$ is given by
Eq.(\ref{e3}):
\beqn
\label{exc}
\Delta E(Q)=4\sin x \cdot\left\{
\ba{ll}
|\sin(x-k_F)|, & 0\!<\!x\!<\!\pi/2\\
|\sin(x+k_F)|, & \pi/2\!<\!x\!<\!\pi
\ea
\right.; \quad x=\frac {|Q|}2\left(1+\frac{k_F}{\pi}\right)
\eeqn
which is shown for $k_F\!<\!\pi/2$ in fig.\ {\ref{fig0}. If
  $k_F\!>\!\pi/2$ one has to substitute $\pi-k_F$ for $k_F$.
The  fermion-hole excitation plot is symmetric when a momentum transfer,
$Q$, changes from 0 to 
$2k_F(1+k_F/\pi)^{-1}$. It formally differs of the results
predicted by Eqs.(\ref{e1}) and (\ref{e2}). These occur due to 
a contribution of order $1/L$ in the first two terms 
in the r.h.s. of Eq.(\ref{en_exc}). However, this contribution is still
important when the finite-size systems are analyzed.

This is noteworthy, that the contraction of the wavevector, 
$2k_F\rightarrow 2k_F(1+k_F/\pi)^{-1}$, when a fermion 
is transfered from one Fermi-point to another,  
 can be attributed to strong ``anti''-correlated properties of our spinless 
``free'' fermions, that forces for using a fictitious lattice description.

As has been shown in \cite{vill} the FP should 
exhibit a power-like decay of correlation functions, e.g.,
\be
\langle\sigma_i^z\sigma_{i+r}^z\rangle\propto r^{-\rho}\cos \pi qx
\label{corrfp}
\ee
with $\rho=(1-q)^2/2$.  This $q$-dependent $\rho$ is also a result of
the fictitious lattice contraction which increases with $q$.  However,
the FP cannot cover the whole range $0<\!q\!<1/2$: it appears to be
unstable at smaller $q$'s and transforms into the PM state through the
mechanism of the Kosterlitz-Thouless transition at $\rho\approx 1/4$
\cite{vill}, corresponding to a wavevector
\be
q_{\rm PM-FP} = 1-1/\sqrt{2}\approx0.292
\label{qpmfp}
\ee

Hamiltonian (\ref{red_1}) allows to go beyond the first order in
$\Gamma$ which has been discussed above.  A non-trivial contribution
in the second order in $\Gamma$ to perturbation theory results in
appearance the terms, which create (annihilate) DW's on next nearest
sites. In fact, a couple of DW's on the sites, say, $j$ and $j+2$, can
be created as a two-step process: First, if the high-energy excitation
is virtually created by $\tau^+_j\tau^+_{j+1}$, then a many-fold
degenerated ground state can be restored by
$\tau^-_{j+1}\tau^+_{j+2}$. The second possibility is in a sequential
process: $\tau^+_{j+1}\tau^+_{j+2}$ and $\tau^-_{j+1}\tau^+_{j}$.
Perturbation theory in the $\Gamma^2$-order also extends the DW
hopping terms up to next nearest neighbors, but this extension is out
of relevance.

Thus, in addition to Hamiltonian (\ref{eff_1t}) we obtain
\beqn
\label{eff_2t}
{\cal H}_{{\rm eff}}^{(2)}=-\gamma\left(\sum_{j=1}^{L-2}
\tau^+_j\tau^+_{j+2}+ 
\tau^+_1\tau^+_{L-1} +\tau^+_2\tau^+_{L}+{\rm HC}\right)
\eeqn
which can be also written in terms of fermions:
\be
\label{eff_2c}
{\cal H}_{{\rm eff}}^{(2)}=-\gamma\left(\sum_{j=1}^{L-2}
c_j^{\dag}c_{j+2}^{\dag}+c_1^{\dag}c_{L-1}^{\dag}+c_2^{\dag}c_{L}^{\dag}+
{\rm HC}\right)
\ee
We put $\gamma=2\Gamma^2/V$ in Eqs.(\ref{eff_2t})-(\ref{eff_2c}) and 
HC denotes Hermitian conjugate.

In spite of a smallness of ${\cal H}_{{\rm eff}}^{(2)}$ as compared to
${\cal H}_{{\rm eff}}^{(1)}$ the former significantly influences on 
the critical properties of the FP: ${\cal H}_{{\rm eff}}^{(2)}$ forces the 
FM--FP transition to be of the Ising type and trasforms this, in fact, into
the FM--PM. This results in a gap opening in the PM phase. 
The excitation spectrum in
a low DW density limit behaves as ($k$ here is the wavevector)
\be
\label{exc2}
\sqrt{(\epsilon-2\Gamma\cos k)^2+(2\gamma\sin k)^2}
\ee
which remains of a single-minimum kind within a narrow interval 
$2\Gamma<\epsilon<2\Gamma-2\gamma^2/\Gamma$, then it develops 
in a double-minimum curve. The gap value, in general, is $\propto \Gamma^2$

The situation on the FP--$<\!2\!>$ boundary is different: This 
state is characterized by a regular DW structure and elementary excitations
driven by ${\cal H}_{{\rm eff}}^{(2)}$ are the four -``leg'' dislocations,
which must be irrelevant \cite{copper}. This ``leg''- number can be easily
illustrated in terms of the ${\cal A}$ and ${\cal B}$ particles.
The FM vacuum  is unstable at the
FM--PM transition with respect to ${\cal H}_{{\rm eff}}^{(2)}$ 
which transforms that vacuum state 
$\dots{\cal B}{\cal B}{\cal B}{\cal B}\dots$ into the set of 
$\dots{\cal B}\dots{\cal B}{\cal A}{\cal A}{\cal B}\dots{\cal B}\dots$
functions, that means $p=2$. On the contrary, the $<\!2\!>$ phase
is described as the $\dots{\cal A}{\cal A}{\cal A}{\cal A}\dots$ vacuum.
According to the definition when ${\cal A}$ (or DW) disappears, it creates
a couple ${\cal B}{\cal B}$. Hence, the elementary excitation due to
${\cal H}_{{\rm eff}}^{(2)}$ can be represented as 
$\dots{\cal A}\dots{\cal A}{\cal B}{\cal B}{\cal B}{\cal B}{\cal A}\dots{\cal A}\dots$, which is irrelevant ($p=4$).

A fictitious lattice as it has been introduced in \cite{vill} is better 
realized in the case of infinite chain with 
no periodic boundary conditions imposed. The fictitious lattice sites are 
enumerated as $\widetilde{m}=m-\eta(m)$, where $\eta(m)=\sum_{i<m}n_i$
is the total number of
fermions from the left of the $m$-th site of a real lattice. 
For fermions on the fictitious lattice we accept the notation
$\widetilde{c}_{\widetilde{j}}$ ($\widetilde{c}_{\widetilde{m}}^{\dag}$).
${\cal H}_{{\rm eff}}^{(1)}$ is formally unchanged:
\be
\label{eff_1cf}
\widetilde{{\cal H}}_{{\rm eff}}^{(1)}=\sum_{\widetilde{m}}\epsilon 
\widetilde{n}_{\widetilde{m}} -
\Gamma(\widetilde{c}^{\dag}_{\widetilde{m}}\widetilde{c}_{\widetilde{m}+1}+
\widetilde{c}^{\dag}_{\widetilde{m}+1}\widetilde{c}_{\widetilde{m}})
\ee
while ${\cal H}_{{\rm eff}}^{(2)}$ takes a form:
\be
\label{eff_2cf}
\widetilde{{\cal H}}_{{\rm eff}}^{(2)}=-\gamma\sum_{\widetilde{m}}
\left(\widetilde{c}^{\dag}_{\widetilde{m}}
\widetilde{c}^{\dag}_{\widetilde{m}+1}P_+(\widetilde{m}+1)+
\widetilde{c}_{\widetilde{m}+1}\widetilde{c}_{\widetilde{m}}
P_-(\widetilde{m}+1)\right)
\ee
where the operator $P_+(\widetilde{m})$ ($P_-(\widetilde{m})$) translates 
all the fermions from the right of $\widetilde{m}$ by two sites to the right 
(left).

\section{Exact diagonalization of finite systems}

The starting point for our numerical investigation of the ground-state
properties of the ANNNI-model (1) is the effective Hamiltonian ${\cal
  H}_{{\rm eff}}^{(1)}$ without creation/annihilation of domain walls
(\ref{eff_1t}), which we call the {\it free model} and the effective
Hamiltonian ${\cal H}_{{\rm eff}}^{(1)}+{\cal H}_{{\rm eff}}^{(2)}$,
which includes creation/annihilation of domain walls (\ref{eff_2t})
and which we call the {\it complete model}. The Hilbert space for
these models consists of all configurations of the original
ANNNI-model, which obey the constraint discussed in the last section.
This constraint reduces the dimension of the Hilbert space
considerably so that much larger system sizes can be diagonalized
(even if one uses the translational and spin flip symmetry of the
original ANNNI Hamiltonian to block diagonalize it first).

\subsection{Methodology}

We reformulate the constraint in such a way that it becomes
suitable for a numerical implementation. The domain walls can be
identified with particles being able to hop to the left or right
provided the constraint will not be violated by this move (moreover at
most one particle can occupy a single site in the dual lattice). Since
a domain wall at bond $j$ comes always with bond $j+1$ free of domain
walls we call this combined object an ${\cal A}$-particle situated at
bond $j$. If no domain wall is at bond $j$ we call it a ${\cal
  B}$-particle provided no other domain wall occurs at bond $j-1$. 
In an obvious notation the following particle
configuration, domain wall (i.e.\ $\tau$-) configuration and spin
configuration (in the $\sigma_z$-representation) correspond to each other
\be
\vert{\cal A}{\cal A}{\cal B}{\cal B}{\cal A}\cdots\rangle
\quad=\quad
\vert10100010\cdots\rangle
\quad=\quad
\left\{
\begin{array}{c}
\mid\uparrow\downarrow\downarrow\uparrow\uparrow\uparrow\uparrow\downarrow\downarrow\cdots\rangle\\
\mid\downarrow\uparrow\uparrow\downarrow\downarrow\downarrow\downarrow\uparrow\uparrow\cdots\rangle
\end{array}
\right.
\ee
Moving a domain wall from bond $i$ to $i\pm1$ means moving an ${\cal
  A}$-particle from site $i$ to $i\pm1$, which is only possible, if at
site $i\pm1$ is a ${\cal B}$-particle.
\be\ba{rcl}
\tau_1^+\tau_2^-\vert{\cal B}{\cal A}\cdots\rangle
&\quad=\quad&\vert{\cal A}{\cal B}\cdots\rangle\\
\tau_1^-\tau_2^+\vert{\cal A}{\cal B}\cdots\rangle
&\quad=\quad&\vert{\cal B}{\cal A}\cdots\rangle
\ea\label{hopping}
\ee
Thus in the particle
representaion the above mentioned constraint is already contained.
Analogous remarks hold for the creation/annihilation of domain walls
described by ${\cal H}_{{\rm eff}}^{(2)}$: in the particle formulation
that means that ${\cal A}$-particles can be created in pairs in
place of four consecutive ${\cal B}$-particles. For instance 
\be\ba{lcl}
\tau_1^+\tau_3^+\vert{\cal B}{\cal B}{\cal B}{\cal B}\cdots\rangle
&\quad=\quad&\vert{\cal A}{\cal A}\cdots\rangle\\
\tau_1^-\tau_3^-\vert{\cal A}{\cal A}\cdots\rangle
&\quad=\quad&\vert{\cal B}{\cal B}{\cal B}{\cal B}\cdots\rangle
\ea\label{creation}\ee
Because of the periodic boundary conditions we have to discriminate
between the cases with and without a domain wall at bond $L$. 
We denote the first group of states with a prime, for instance 
(in the case of L=8):
\be 
\vert{\cal A}{\cal B}{\cal B}{\cal A}{\cal A}\rangle'
\quad=\quad\vert01000101\rangle
\label{primed}
\ee
(note that in this notation the rightmost particle is always an 
${\cal A}$-particle), whereas those without prime denote states
without domain wall between $L$ and $1$, e.g.:
\be
\vert{\cal A}{\cal B}{\cal B}{\cal A}{\cal A}\rangle
\quad=\quad\vert10001010\rangle
\label{unprimed}
\ee 
This state is simply a circular left shift of the primed state in
(\ref{primed}), in fact for each primed state there is a unique
unprimed state that can be obtained from the former via a circular
left shift. However, there are of course more primed states than
unprimed ones.

Thus we consider different primed and unprimed subspaces characterized
by the number of ${\cal A}$-particles (i.e.\ number of domain walls),
which is conserved under the action of ${\cal H}_{{\rm eff}}^{(1)}$
(note, however, that the latter mixes states of the primed and unprimed
subspaces by moving domain walls to or from the bond $L$. The
dimension of these subspaces is simply given by the number of
different possibilities to distribute $n_A$ and $n_A-1$ particles on
$L-n_A$ and $L-n_A-1$ sites, respectively.
\be\ba{rlc}
{\rm dim}_{n_A}&=&\binomial{L-n_A}{n_A}\;,\\
{\rm dim}_{n_A}'&=&\binomial{L-n_A-1}{n_A-1}\;.
\ea\ee
The dimension of the whole Hilbert space we are considering is
\be
{\rm dim}_{{\cal H}_{eff}}=1+\sum_{n_A=1}^{L/2}\left\{
\binomial{L-n_A}{n_A}+\binomial{L-n_A-1}{n_A-1}\right\}
\ee
which is a much smaller number than the dimension of the original
Hilbert space, which is $2^L$.

\bc
\bt{|c||l|l|l|} \hline
 L & dim$_H$  & 2$^L$       & dim$_H$ / 2$^L$ \\ \hline
 8 & 23     & 256        & 8.984375e-02 \\
12 & 162    & 4096       & 3.955078e-02 \\
16 & 1103   & 65536      & 1.683044e-02 \\
20 & 7563   & 1048576    & 7.212639e-03 \\
24 & 51842  & 16777216   & 3.090024e-03 \\
28 & 355323 & 268435456  & 1.323681e-03 \\
32 & 2435423& 4294967296 & 5.670411e-04 \\ \hline
\et
\vskip0.5cm

{\baselineskip13pt

{\bf Table I}: Dimensionality of the reduced Hilbert space of our
effective Hamiltonians (second column) and the dimensionality of the
original Hilbert space (third column).}
\ec
\vskip0.3cm

From table I it becomes obvious that the storage requirements for
diagonalizing the effective Hamiltonian is significantly smaller. This
is still the case if one uses all symmetries of the original ANNNI
Hamiltonian to blochdiagonalize it first (by which can reduce the
storage requirement by roughly a factor $1/L$). Note that in the
FP-phase it is not a priori clear in which wave number sector of the
original Hilbert state the ground state is situated. Hence all of them
have to be considered, as has been done in \cite{duxbury} for lattice
site up to $L=16$. We were able to diagonalize easily L=32 systems on
workstations with a reasonable RAM without reading or writing to the
hard disk.
 
In order to enumerate the states within the subspaces efficiently we
recur to a scheme that is frequently used in the context of quantum
spin chains confined to subspaces with constant magnetization
\cite{Hirsch}. Let us fix $L$ to be a multiple of $4$ (so that the
periodic boundary conditions are fully compatible with the ground
state in the anti-phase $<\!2\!>$). Let $\psi_{n_A}(n)$ be an unprimed state
with $n_A$ ${\cal A}$-particles: 
\be
\psi_{n_A}(n)=\vert{\cal A}(p_1)\cdots{\cal A}(p_{n_A})\rangle
=\vert\underbrace{{\cal B}\cdots{\cal B}}_{p_1-1}{\cal A}
\underbrace{{\cal B}\cdots{\cal B}}_{p_2-p_1-1}{\cal A}\cdots\cdots\rangle\;,
%\quad n\in\{0,\ldots,{\rm dim}_{n_A}-1\}\;,
\label{notation_def}
\ee
where $p_i\in\{1,\ldots,L-n_A\}$ denotes the position of the $i$-th
${\cal A}$-particle (counted from the left). Then the following definition
yields a one-to-one correspondence between the possible configurations
and a number $n\in\{0,\ldots,{\rm dim}_{n_A}-1\}$ :
\be
n=\sum_{i=1}^{n_A}
\binomial{p_i-1}{i}\qquad{\rm with}\quad\binomial{i-1}{i}=0\;.
\ee
The same definition is used for primed and unprimed states, which we
denote by $\psi_{n_A}(n)$ and $\psi_{n_A}'(n)$, respectively. The
Lanczos routine we use to calculate the ground state and first excited
state generates the Hamiltonian each time it is needed, for this
reason we need to know how the various hopping, creation and
annihilation operators act upon the basis states we have chosen. This
can either be done by a hashing technique, which is used frequently
for  arbitrary quantum spin chains, or by explicitely calculating the
number of the transformed state if possible. Fortunately in our case
the latter is straightforward, and in Appendix D we list all relevant
formulas that we need to generate the non-zero matrix elements for
${\cal H}_{{\rm eff}}^{(1)}$ and ${\cal H}_{{\rm eff}}^{(2)}$.

With the help of nowadays standard Lanczos routines we calculated the
ground state and the first excited state of systems of size up to
L=32. The effective Hamiltonians we derived are expected to be good
approximations to the original ANNNI model for small values of
$\Gamma/J$. We confined ourselves to the values $\Gamma/J=0.05$,
$0.1$, $0.2$ and $0.5$, where we calculated all physical quantities of
interest for $\kappa$-values in intervals of $5\cdot10^{-4}$. We
performed extensive checks of our code by comparing our numerical
estimates for the ground state energy and the gap with the exactly
known values for the free model (\ref{eff_1t}). We also compared our
results obtained for ${\cal H}_{\rm eff}^{(1)} + {\cal H}_{\rm eff}^{(2)}$ at
small values of $\Gamma/J$ with those for the original ANNNI model and
found no significant deviations.

\vskip2cm

\subsection{Results}

In order to check the quality of the free fermion description we first
calculated the ground states of the original ANNNI model
(\ref{ham})-(\ref{mast_eq}) for a modest system size (up to L=20). The
``classical'' energy of a state
\be
E_{\rm class}=\langle\psi\vert H_{\rm cl}\vert\psi\rangle
\ee
which is $E_{\rm class}=-{\cal J}(\sum_i S_i S_{i+1} - \kappa\sum_i
S_i S_{i+2})$ for an eigenstate of the $\sigma_i^z$ operators contains
the information on the average number of domain walls (or ${\cal
A}$-particles) in the state $\psi$, if this last does allow neither 
$\,\dots\uparrow\downarrow\uparrow\dots\,$, nor 
$\,\dots\downarrow\uparrow\downarrow\dots\,$ to appear. 
With such a constraint imposed on a state with exactly
$n_A\in\{0,2,4,\ldots,L/2\}$ domain walls, each of them costing an energy
$\epsilon=-(4\kappa-2){\cal J}$ with respect to the FM
state, we have
\be
E_{\rm class}/{\cal J} + (1-\kappa) L = n_A\cdot(4\kappa-2)
\ee
Thus the comparison of the l.h.s.\, which we call $DE$, with the set
of straight lines provides a measure of the average number of ${\cal
A}$-particles plus an indication of the appropriateness of the free
fermion concept in this context. In fig.\ \ref{fig1} we show the
result for various values of $\Gamma$ and we see that the smaller
$\Gamma$ the better the agreement of various parts of the
$DE$-curves.  Furthermore, for increasing $\Gamma$ the ground state is
more a superposition of various particle eigenstates {\it in the
vicinity of the FM-PM transition}.

Now we turn our attention to the effective Hamiltonians. The average
number of domain walls, simply given by $\langle n_A\rangle =
\langle\psi\vert \sum_i \tau_i^+\tau_i^-\vert\psi\rangle$ is calculated
for $\psi$'s of the ground state and of the first excited state 
of the complete model (\ref{eff_1t})+(\ref{eff_2t}). 
For $\Gamma$ not too large one observes well defined
regions with constant value for $n_A$ in the ground state and the
first excited state. See fig.\ \ref{fig2} for an example. The number
of ${\cal A}$-particles increases monotonically (in steps of 2) for
increasing $\kappa$. The points where the particle number jumps are of
special interest. The particle number of the first excited state
$\langle n_A\rangle _1$ jumps first abruptly 
(approaching the transition points from
either side) and then it changes roles with the ground state. As a
consequence the gap (i.e. the energy difference between ground state
and first excited state) gets very small here. 

Note that in the FM-phase the gap does {\it not} vanish exponentially
with system size for the effective Hamiltonians, because in the
particle representation the two degenerated states with all spins up or
all spins down are represented as {\it one} state. Therefore the gap
closes in the infinite system only {\it at} the FM-PM transition. On
the other hand the gap stays zero (or exponentially small for finite
sizes) throughout the $<\!2\!>$ phase also for the effective Hamiltonians:
The 4-fold degeneracy there is only reduced by a factor two via
the elimination of the spin-flip symmetry and a degeneracy between
corresponding primed and unprimed states is left.

As a significant difference of the complete model with respect to the
free model we note that at the special $\kappa$-values, where the
particle number changes and which can be calculated exactly via the
formula (\ref{gse}), the gap of the free model (\ref{eff_1t})
closes completely, i.e.\ $\Delta E_{n_A\to n_A\pm2}^{\rm free}=0$.
In the complete model the gap-value on the boundary, say, $n_A\to n_A+2$, 
can be easily 
estimated if we confine our consideration with these two competing states
only, that results in
\be
\label{gap_plus2}
\Delta E_{n_A\to n_A+2}=2\langle \psi_{n_A+2}
\vert {\cal H}_{\rm eff}^{(2)}\vert\psi_{n_A}\rangle
\ee
These special gap-values increase with
$\Gamma$ because of the $\Gamma$-dependent ${\cal H}_{\rm eff}^{(2)}$.
In particular more pronounced is the gap increasing for the lower
$\kappa$-values. In this range the boundary $n_A\to n_A+2$
cannot be considered as well isolated from other ``neighboring'' states,
{\it i.e.}, $n_A-2$, $n_A+4$, etc.
This observation will turn out as a hint for the
existence of the PM-FP transition.  For higher values of $\Gamma/J$
the resulting picture is therefore slightly different: the transitions
smear out and $\langle n_A\rangle(\kappa)$ and $\Delta E$ become smoother as 
can be
seen in fig.\ \ref{fig3}. 

Let us interpret this picture {\it cum grano salis}:
One might tentatively locate the FM-PM transition (for $\Gamma/J=1/5$)
roughly at $\kappa\approx0.41$, where the gap should approach zero
like $\Delta E\sim 1/L$ (since the transition is expected to be in the
Ising universality class, where $\nu=z=1$). Between $\kappa=0.41$ and
let us say $\kappa\approx0.48$ the average particle number increases
from zero to 8 (for L=24), but the individual transitions observable
in fig.\ \ref{fig2} {\it melt} together to form a rugged plateau. Only when
$n_A$ gets larger than some value (whose significance we will clarify
later) the gaps at the individual transitions try to close again.
With increasing system size these gaps (for $\kappa$ larger than
roughly 0.48) melt together, too, but in the limit $L\to\infty$ they
will form a curve $\Delta(\kappa)=0$ in this region, which is simply
the gapless floating phase with a {\it quasi}-long range, i.e.,
algebraically decaying spin correlations.

It is obvious that in order to observe this scenario in its pure form
one has to go to enormous system sizes, which is not feasible yet.
Nevertheless we can clearly demonstrate the qualitative difference
between the PM phase and the FP phase by explicitly studying the
spin correlations in both regions of the phase diagram for intermediate
system sizes.

The spin correlation function is defined via
\be
\ba{rcl}
C(r)&=&\langle\psi\vert\sigma_i^z\sigma_{i+r}^z\vert\psi\rangle\\
    &=&\ds\sum_{n_A=0}^{L/2}\sum_{n=0}^{{\rm dim}_{n_A}}
       \psi_{n_A}^2(n)\cdot\frac{1}{L}\sum_{i=1}^L
        S_i[\psi_{n_A}(n)]\,S_{i+r}[\psi_{n_A}(n)]
       \;+\;{\rm primed\;states}
\ea
\label{corr}
\ee
Here $\underline{S}[\psi_{n_A}(n)]$ ($\underline{S}=S_1,\ldots,S_L$)
means the spin configuration that is equivalent to the state number
$n$ with $n_A$ particles (since there are always two of them we choose
the one with the first spin up $S_1=+1$). Of course we have to
initialize such a mapping in our program once, afterwards this table can be
used whenever correlations have to be calculated.

First we take a look at the structure function since this directly
relates to the particle number $n_A$ discussed above. We define it as follows:
via
\be
f_q(\kappa)=\frac{1}{L^2}\sum_{i,j} \,\cos\,\left(q(i-j)\right)\,
\langle\psi\vert\sigma_i^z\sigma_j^z\vert\psi\rangle
\label{struc}
\ee
for $q= n / L$ ($n=0,1,\ldots,L-1$) so that
\be
C(r)=\sum_q \cos(qr) f_q
\ee
In fig.\ \ref{fig4a} we show the result for $f_q(\kappa)$ for
$\Gamma/J=1/20$ in comparative plots for the free and the complete
model. For fixed wave number $q$ the structure function $f_q(\kappa)$
is simply a step function for the free model, the plateaus located at
the $\kappa$-intervals with constant particle number $n_A$. For fixed
$\kappa$ the wave number $q_{\rm max}$ with the maximum amplitude
$f_{q_{\rm max}}(\kappa)>f_q(\kappa)$ for $q\ne q_{\rm max}$ 
is related to the particle number $n_A$ via
\be
q_{\rm max}=n_A/2L
\ee
(note that $n_A$ is a good quantum number in the free model). For
small $\Gamma/J$ we observe that the steps get rounded, but nothing
dramatic (for these system sizes) happens. If we increase $\Gamma/J$,
as is done in fig.\ \ref{fig4b}, the neighboring plateaux begin to mix
(to melt, see above). They also shift their location to larger values
of $\kappa$. However, these changes seem to become less significant as
soon as $q_{\rm max}$ is larger than $1/4$ (actually, as stated in the
last section in Eq.(\ref{qpmfp}), the theoretical estimate for this 
relevance-threshold is approximately $q=0.292>1/4$). 

This statement finds its strongest support by looking at the spin
correlation function (\ref{corr}), shown in fig.\ \ref{fig5} and
\ref{fig6} for L=32, directly.  For the free model of finite size
$L$ with periodic boundary conditions one would according to
(\ref{corrfp}) expect that
\be
G(r)=a\,\cos(r\pi q_{\rm max})\cdot\{ r^{-\rho} + (L-r)^{-\rho} \}
\quad{\rm with}\quad\rho=\frac{1}{2}(1-q_{\rm max})^2\;,
\label{crfit}
\ee
where $a$ is a fit
parameter and $q_{\rm max}$ has to be determined from the structure
function. In order to resolve the correlations over as large as
possible distances we took here the largest possible system sizes.
For L=32, however, we had to confine ourselves to a smaller number of
parameter values. In the case $\Gamma/J=1/20$, in which the
differences between the complete and the free model are not too large,
we took simply the middle of the plateaus of the structure function,
whereas for $\Gamma/J$ we took the location of the maxima of
$f_q(\kappa)$ shown in fig.\ \ref{fig7}.

What is shown in fig.\ \ref{fig5} and \ref{fig6} is a comparative plot
of $C(r)$ and $G(r)$ with $a=\sqrt{\sum_r C^2(r) / \sum_r
\tilde{G}^2(r)}$, $\tilde{G}(r)=G(r)/a$. We observe that for 
$q_{\rm max}>1/4$ the free model correlation
function $G(r)$ fits $C(r)$ in an excellent way. For $q_{\rm
max}\le1/4$, however, one recognizes sigificant differences between
$G(r)$ and $C(r)$, most dramatic for the smallest wave numbers, which
mean closest to the FM-PM transition of the complete model. The
$q_{\rm max}=1/12$ and $q_{\rm max}=1/6$ curves for $\Gamma/J=1/5$
definitely decay faster than algebraic. We obtained an excellent fit
(shown in fig.\ \ref{fig8}) for the $q_{\rm max}=1/16$ curve by a 
superpostion of an exponentially damped $q=0$ and $q=1/16$ oscillation:
\be
C_{\kappa=0.4085;\Gamma/J=5}(r)=
(1-a+a\cos(r\pi/16))\cdot(e^{-r/\xi}+e^{-(32-r)/\xi})
\label{fit}
\ee
with $a=0.21$ and $\xi=3.6$. For us this is ample evidence that for
fixed $\Gamma/J$ one enters first a paramagnetic phase with
exponentially decaying correlations by increasing $\kappa$ from the
FM-phase. Only when $q_{\rm max}$ gets larger then the above
mentioned value, one enters the FP-phase, which for the free model
extends over the whole region between the $<\!2\!>$  and FM phases.

\section{Summary}

Below we summarize a few important points of this work.

\begin{itemize}

\item Instead of the original 1$d$ ANNNI model in a transverse field
  we consider a reduced model which we show to be a very reasonable
  modification when the competition parameter $\vert\kappa -1/2\vert$
  as well as the quantum parameter $\Gamma/{\cal J}$ are small.
\item The effective Hamiltonian ${\cal H}_{\rm eff}^{(1)}+{\cal
    H}_{\rm eff}^{(2)}$ is most easily visualized with the ${\cal A}-{\cal
    B}$ particle representation.  The latter has been adapted to the
  periodical boudary conditions and efficiently used in
  numerical calculations.
\item Analytical calculations here were done not for re-discovering after
  \cite{vill} the essential physics, accompanying the scheme of phase
  transitions FM-PM-FP-$<\!2\!>$  in infinitely long chains. Most analytical
  results are derived for further applications in a numerical scheme.
\item The exact numerical diagonalization technique (Lanzcos algorithm)
  was used on the restricted basis states, all of them have been
  enumerated, as well as all the non-zero matrix elements generated by
  Hamiltonian of the complete model were stored. Along this line we
  perform calculations for systems of size up to L=32 with standard
  work-stations.
\item The set of Figs. 5-10 convincingly illustrates a different origin
  of the critical behavior at larger and smaller $\kappa$'s. Most
  likely, the PM-FP transition takes place close to or even at a
  modulation $q$ given by the theory of the standard
  Kosterlitz-Thouless-like transition.

\end{itemize}

\acknowledgments
This work has been performed within the Sonderforschungsbereich (SFB)
341 K\"oln-Aachen-J\"ulich. G.\ U.\ thanks the Institut f.\
Theoretische Physik of the University of K\"oln for its kind
hospitality and H.\ R.\ thanks S.\ Dasgupta for numerous fruitful
discussions and numerical experiments on the ANNNI-model.

\vfill
\eject

\subsection*{Appendix A}
\vspace{-2mm}

\renewcommand{\theequation}{A.\arabic{equation}}
\setcounter{equation}{0}
\begin{appendix}
We treat $\tau^z_j$ as a domain wall operator, defined on the lattice site $j$
of the dual lattice. It takes two values, $\pm 1$:
$+1$ ($-1$) signals of absence (presence) of a domain wall. Mathematically, 
this can be expressed by 
\be
\label{t_z}
\tau^z_j=\sigma_{j-1/2}^z\sigma_{j+1/2}^z. 
\ee
Because of the periodic boundary conditions we also define 
\be
%\label{boun_t_z}\nonumber
\tau^z_1=\sigma_{1/2}^z\sigma_{3/2}^z \quad {\rm and}\quad 
\tau^z_L=\sigma_{1/2}^z\sigma_{L-1/2}^z. 
\ee
Also evidently, that
\be
\label{tt_z}
\tau^z_j\tau^z_{j+1}=\sigma_{j-1/2}^z\sigma_{j+3/2}^z. 
\ee
For $x$-components we accept the following definitions: 
\be
\label{s_x}
\sigma_{j+1/2}^x=\tau^x_j\tau^x_{j+1}
\ee
 and
\be
%\label{boun_s_x}\nonumber
\sigma_{1/2}^x=\tau^x_1\tau^x_L \quad {\rm and}\quad 
\sigma_{L-1/2}^x=\tau^x_{L-1}\tau^x_L.
\ee
With using Eqs.(\ref{t_z}),(\ref{tt_z}) and (\ref{s_x}) we can easily obtain 
the form of Eq.(\ref{red_1}) from (\ref{ham}).

This transformation set allows to calculate the correlation functions in 
terms of $\tau$'s, for example:
\be
\label{cor_zz}
\langle\sigma_{r_1-1/2}^z\sigma_{r_2+1/2}^z\rangle=\langle\prod_{j=r_1}^{r_2}
\tau^z_j\rangle
\ee
\end{appendix}
\vspace{-8mm}

\subsection*{Appendix B}
\vspace{-2mm}

\renewcommand{\theequation}{B.\arabic{equation}}
\setcounter{equation}{0}
\begin{appendix}
In Eq.(\ref{wf1}) $m$'s are supposed to be 
arranged in order, satisfying the following constraints: 
$m_1<m_2-1$, $m_2<m_3-1,\dots,m_{k-1}<m_k-1$, $m_k<m_1+L-1$. 
We transfer these constraints on the amplitudes $f$: 
\be
\label{constr_1}
f(\dots,m,m+1\dots)=0
\ee
\be
\label{constr_2}
f(1,\dots,L)=0.
\ee
In spite of the irregular $1\leftrightarrow L$ hopping term in 
Eq.(\ref{eff_1c}) the eigen amplitudes $f$ satisfy the regular equations:
\beqn
\nonumber
(E-k\epsilon)f(m_1,m_2,\dots,m_k)=
-\Gamma \sum_{a=\pm 1}(f(m_1+a,m_2,\dots,m_k)+\\
f(m_1,m_2+a,\dots,m_k)+\dots +f(m_1,m_2,\dots,m_k+a)).
\label{eig_eq}
\eeqn
Using a Bethe substitution for $f$'s (see Eq.(\ref{bethe})) we obtain 
\be
\label{ener}
E=k\epsilon-2\Gamma \sum_{j=1}^k\cos q_j.
\ee
Eqs.(\ref{constr_1}) and (\ref{constr_2}) yield
\be
\label{constr_1'}
\xi_{\dots ij\dots}e^{\imath q_j}+\xi_{\dots ji\dots}e^{\imath q_i}=0
\ee
and 
\be
\label{constr_2'}
\xi_{i\dots j}e^{\imath (q_i+Lq_j)}+\xi_{j\dots i}e^{\imath (q_j+Lq_i)}=0,
\ee
respectively. In turn, from Eqs.(\ref{constr_1'}) and (\ref{constr_2'}) 
one can arrive to
\be
\label{qij}
q_i-q_j=\frac {2\pi}{L-k}n_{ij}
\ee
where $n_{ij}$ are integer numbers. Additional equations imposed on $q$'s
may be obtained from Eq.(\ref{eig_eq}) at $m=1$. Formally, it is equivalent
to $f(0,m_2,\dots,m_k)=f(m_2,\dots,m_k,L)$, that is
\be
\label{constr_3'}
\xi_{ijs\dots t}=\xi_{js\dots ti}e^{\imath Lq_i}
\ee
which after a simple algebra results in
\be
\label{qi}
q_i=\frac{\pi n^{\rm odd}_i }{L-k}-\frac 1{L-k}\sum_{j=1}^kq_j,
\ee
$\{n^{\rm odd}\}$, the set of odd integer numbers.

The content of this Appendix may be summarized in this last Eq.(\ref{qi})
and the form of the the eigenfunctions (\ref{wf}).
\end{appendix}
\vspace{-8mm}

\subsection*{Appendix C}
\vspace{-2mm}

\renewcommand{\theequation}{C.\arabic{equation}}
\setcounter{equation}{0}
\begin{appendix}
For the ground state of $k$ DW's the set $\{n^{\rm odd}\}$ in Eq.(\ref{qi}) 
counts all the equidistant odd numbers from $-k+1$ to $k-1$. Let us take away 
one number from that set, say, $n^{\rm odd}_-\!=2\nu_-\!-1$, with 
$-k/2+1\!<\!\nu_-\!<\!k/2-1$,
and add another number $n^{\rm odd}_+\!=2\nu_+\!-1$, with $\nu_+\!>\!k/2$. 
With this definition
$\sum n^{\rm odd}_i=2(\nu_+-\nu_-)$ and after summation in both parts of 
(\ref{qi}) we get
\be
\label{sum_qi}
Q=\sum_i q_i=\frac {2\pi(\nu_+-\nu_-)}L
\ee
The wavevectors in Eq.(\ref{qi}) are now determined and the energy of such 
an excitation counted from the ground state energy takes a form:
\beqn
\label{en_exc}
\Delta E(Q)=-2\sum_i \cos q_i-E_{\rm gs}=
 -2\!\!\!\sum_{\nu=-k/2+1}^{k/2}\!\!\!\cos\frac {(2\nu -1)\pi-Q}{L-k}-
\nonumber
\\
-2\cos\frac {(2\nu_+\!-1)\pi-Q}{L-k}
+2\cos\frac {(2\nu_-\!-1)\pi-Q}{L-k}
+2\!\!\!\sum_{\nu=-k/2+1}^{k/2}\!\!\!\cos\frac {\pi(2\nu-1)}{L-k}=
\nonumber
\\
=2\left(1-\cos\frac{Q}{L\!-\!k}\right)\sin\frac{\pi k}{L\!-\!k}
\left/
\sin\frac{\pi}{L\!-\!k}\right.- 
\nonumber\\
-2\sin\frac{Q}{L\!-\!k}\left[\sin\frac{\pi(2\nu_+\!-1)}{L\!-\!k}
-\sin\frac{\pi(2\nu_-\!-1)}{L\!-\!k}\right] -
\nonumber
\\
-2\cos\frac{Q}{L-k}\left[
\cos\frac{\pi(2\nu_+\!-1)}{L-k}-\cos\frac{\pi(2\nu_-\!-1)}{L-k}\right]
\eeqn

Using Eq.(\ref{en_exc}) we can analyse the lower lying excitations
\begin{enumerate}
\item
$n^{\rm odd}_-\!=k\!-\!1$, $n^{\rm odd}_+\!=k\!+\!1$ ~$\Longrightarrow Q=2\pi/L$.
\newline
The excitation energy with a vanishing momentum transfer $Q$ in the 
$L\rightarrow\infty$ limit reads in the leading order in $1/L$
(this contribution comes from the last term of the r.h.s. of Eq.(\ref{en_exc})):
\be
\label{e1}
\Delta E(Q)\approx\frac{4\pi}{L-k}\sin\frac{\pi k}{L\!-\!k}=
2\Delta q\sin\frac{\pi k}{L\!-\!k}
\ee
where $\Delta q$ is a $q$ spacing (cf Eq.(\ref{qij}))
\item
$n^{\rm odd}_-\!=-k\!+\!1$, $n^{\rm odd}_+\!=k\!+\!1$ ~$\Longrightarrow Q=2\pi k/L$.
\newline
All the terms in the r.h.s. of Eq.(\ref{qij}) contribute in the leading order
in $1/L$. This excitation occurs at the energy 
\be
\label{e2}
\Delta E(Q)\approx 2\Delta q\left(1-\frac kL\right)^2\sin\frac{\pi k}{L\!-\!k}
\ee
which {\it differs} from (\ref{e1}). 
\item
For not a special value of $Q$ a leading contribution, $O(1)$,
comes from the last term of the r.h.s. of Eq.(\ref{en_exc})). 
To leading order it can be written as:
\be
\label{e3}
\Delta E(Q)= 4\sin x \sin (x+2\pi\nu_-/(L-k)),\quad
x=\frac {Q L}{2(L-k)}.
\ee
We used (\ref{sum_qi}) to obtain the form of Eq.(\ref{e3}). Then in order
to select the lowest energy values at fixed $Q$ we consider following
inequalities:
\begin{itemize}
\item
$x\!<\!k_F=\pi k/(L-k)$. The possible range of $\nu_-$'s becomes
$LQ/(2\pi)-k/2\!<\!\nu_-\!<\!k/2$. The minimum of (\ref{e3}) is reached at 
the lower limit, that results in $\Delta E(Q)= 4\sin x \sin (k_F-x)$
\item
$k_F\!<\!x\!<\!\pi-k_F$. Now a possible range of $\nu_-$'s is
$-k/2\!<\!\nu_-\!<\!k/2$. Two extreme possibilities should be checked,
one arrives to the form $\Delta E(Q)= 4\sin x \sin (x-k_F)$, another
results in $\Delta E(Q)= 4\sin x \sin (k_F+x)$. The former realizes
the minimum at $k_F\!<\!x\!<\!\pi/2$, the latter is correct at
$\pi/2\!<\!x\!<\!\pi-k_F$.
\item
$\pi-k_F\!<\!x\!<\!\pi\longrightarrow-k/2\!<\!\nu_-\!<\!L-3k/2-LQ/(2\pi)$.
A true minimum corresponds to $\Delta E(Q)= 4\sin x \sin (-x-k_F)$
\end{itemize}
\end{enumerate}
\end{appendix}
\vfill
\eject

\subsection*{Appendix D}
\vspace{-2mm}

\renewcommand{\theequation}{D.\arabic{equation}}
\setcounter{equation}{0}
\begin{appendix}
  
In this Appendix we list the formulas that determine the action of
various hopping, creation and annihilation operators on states
$\psi_{n_A}(n)$ and $\psi_{n_A}'(n')$. First remember the definition
of our notation (\ref{notation_def}). In what follows the numbers $n$
and $n'$ are always given by

\be
n=\sum_{j=1}^{n_A}\binomial{p_j-1}{j}\quad{\rm and}\quad
n'=\sum_{j=1}^{n_A-1}\binomial{p_j-1}{j}\;,
\ee

\parindent=0cm

First we consider the hopping term occuring in (\ref{eff_1t}):
\medskip

\underline{{\bf Hopping to the right} $\tau_i^-\tau_{i+1}^+$}\\
\medskip
%\nopagebreak
The matrix elements of the operator
$\sum_{i=1}^{L}\tau_i^-\tau_{i+1}^+$ in the particle representation is
non-zero whenever it is possible to move an $\cala$-particle to the
right. This means there has to be a $j\in\{1,\dots,n_A\}$ in such a way that
$p_{j+1}>p_j+1$, compare with (\ref{hopping}). We have to take special
care of the case $i=L-1$ or $L$ (i.e.\ hopping to or from the periodic
boundary), when an umprimed state transforms into a primed state and
vice versa.
\medskip

$i\in\{1,\ldots,L-2\}$, i.e.\ $p_{n_A}<L-n_A$:
%\nopagebreak
\be\ba{rcl}
\psi_{n_A}(n)=\vert\cdots{\cal A}(p_j){\cal B}\cdots\rangle
&\longrightarrow&
\psi_{n_A}(m)=\vert\cdots{\cal B}{\cal A}(p_j+1)\cdots\rangle\\
m&=&n+\binomial{p_j-1}{j-1}
\ea\ee
The same for a primed state $\psi_{n_A}'(n')$.

$i=L-1$, i.e.\ $p_{n_A}=L-n_A$:
%\nopagebreak
\be\ba{rcl}
\psi_{n_A}(n)=\vert{\cal B}\cdots{\cal A}\rangle
&\longrightarrow&\psi_{n_A}'(m')=\vert\cdots{\cal B}{\cal A}\rangle'\\
m'&=&\displaystyle\sum_{j=1}^{n_A-1}\binomial{p_j-2}{j}
\ea\ee

$i=L$, i.e.\ all primed states:
%\nopagebreak
\be\ba{rcl}
\psi_{n_A}'(n)=\vert{\cal B}\cdots{\cal A}\rangle'
&\longrightarrow&\psi_{n_A}(m)=\vert{\cal A}\cdots{\cal B}\rangle\\
m&=&\displaystyle\sum_{j=1}^{n_A-1}\binomial{p_j-1}{j+1}
\ea\ee
\medskip

\underline{{\bf Hopping to the left} $\tau_i^+\tau_{i+1}^-$}\\
\medskip
$i\in\{1,\ldots,L-2\}$, i.e.\ $p_{n_A}<L-n_A$:
%\nopagebreak
\be\ba{rcl}
\psi_{n_A}(n)=\vert\cdots{\cal B}{\cal A}(p_j)\cdots\rangle
&\longrightarrow&\psi_{n_A}(m)=\vert\cdots{\cal A}(p_j-1){\cal B}\cdots\rangle\\
m&=&n-\binomial{p_j-1}{j-1}
\ea\ee
The same for a primed state $\psi_{n_A}'(n')$.

$i=L$, i.e.\ $p_1=1$:
%\nopagebreak
\be\ba{rcl}
\psi_{n_A}(n)=\vert{\cal A}\cdots{\cal B}\rangle
&\longrightarrow&\psi_{n_A}'(m')=\vert{\cal B}\cdots{\cal A}\rangle'\\
m'&=&\displaystyle\sum_{j=2}^{n_A}\binomial{p_j-1}{j-1}
\ea\ee

$i=L-1$, i.e.\ all primed staes:
%\nopagebreak
\be\ba{rcl}
\psi_{n_A}'(n')=\vert\cdots{\cal B}{\cal A}\rangle'
&\longrightarrow&\psi_{n_A}(m)=\vert{\cal B}\cdots{\cal A}\rangle\\
m&=&\binomial{L-n_A-1}{n_A}+\displaystyle\sum_{j=1}^{n_A-1}\binomial{p_j}{j}
\ea\ee

Next we consider the creation and annihilation operators occuring in
the complete model via (\ref{eff_2t}). Again we have to take special
care of the cases in which a domain wall at the bond linking site $1$
and site $L$ is created or annihilated.
\medskip

\underline{{\bf Creation of domain walls} $\tau_i^+\tau_{i+2}^+$}\\
\medskip
%\nopagebreak
The non-zero matrix elements of the operator $\sum_{i=1}^L
\tau_i^+\tau_{i+2}^+$ have to be determined via the rule
(\ref{creation}).  Let us denote with $p_y$ a position between two
succesive $\cala$-particles at positions $p_x$ and $p_{x+1}$ (with
$p_{x+1}-p_x\ge4$) where a new pair of $\cala$-particles can be
created.
\medskip

$p_y\in\{1,\ldots,n_A-3\}$:
%\nopagebreak
\be\ba{crcl}
&\psi_{n_A}(n)&=&\vert\cala(p_1)\cdots\cala(p_x)
\;[\calb(p_y)\calb(p_y+1)\calb(p_y+2)\calb(p_y+3)]\;
\cala(p_{x+1})\cdots\cala(p_{n_A})\rangle\\
\longrightarrow&\psi_{n_A}(m)&=&\vert\cala(p_1)\cdots\cala(p_x)\;
\cala(p_y)\cala(p_y+1)\;
\cala(p_{x+1}-2)\cdots\cala(p_{n_A}-2)\rangle\\
&m&=&\ds\sum_{j=1}^x\binomial{p_j-1}{j}+\binomial{p_y-1}{x+1}
+\binomial{p_y}{x+2}+\sum_{j=x+1}^{n_A}\binomial{p_j-3}{j+2}
\ea\ee
The same for a primed state $\psi_{n_A}'(n')$ with $n_A$ replaced by
$n_A-1$ in the last sum.

$p_y=L-n_A-2$:
%\nopagebreak
\be\ba{crcl}
&\psi_{n_A}(n)&=&\vert\calb\;\cala(p_1)\cdots\cala(p_{n_A})\;
\calb\calb\calb\rangle\\
\longrightarrow&\psi_{n_A}'(m')&=&\vert\cala(p_1-1)\cdots\cala(p_{n_A}-1)
\cala(L-n_A-3)\cala\rangle'\\
&m'&=&\ds\sum_{j=1}^{n_A}\binomial{p_j-2}{j}+\binomial{L-n_A-4}{n_A+1}
\ea\ee

$p_y=L-n_A-1$:
%\nopagebreak
\be\ba{crcl}
&\psi_{n_A}(n)&=&\vert\calb\calb\;\cala(p_1)\cdots\cala(p_{n_A})\;
\calb\calb\rangle\\
\longrightarrow&\psi_{n_A}(m)&=&\vert\cala\;\cala(p_1-1)\cdots\cala(p_{n_A}-1)\;
\cala\rangle\\
&m&=&\ds\sum_{j=1}^{n_A}\binomial{p_j-2}{j+1}+\binomial{L-n_A-3}{n_A+2}
\ea\ee

$p_y=L-n_A$:
%\nopagebreak
\be\ba{crcl}
&\psi_{n_A}(n)&=&\vert\calb\calb\calb\;\cala(p_1)\cdots\cala(p_{n_A})\;
\calb\rangle\\
\longrightarrow&\psi_{n_A}'(m')&=&\vert\cala(1)\;
\cala(p_1-2)\cdots\cala(p_{n_A}-2)\;\cala\rangle'\\
&m'&=&\ds\sum_{j=1}^{n_A}\binomial{p_j-3}{j+1}
\ea\ee
\medskip

\underline{{\bf Annihilation of domain walls} $\tau_i^-\tau_{i+2}^-$}\\
\medskip
%\nopagebreak
Now let there be two successive $\cala$-particles at position
$p_x$ and $p_{x+1}=p_x+1$, so that they can be annihilated by
$\tau_i^-\tau_{i+2}^-$ for some suitably chosen $i$.
\medskip

$p_x\in\{1,\ldots,L-n_A-3\}$:
%\nopagebreak
\be\ba{crcl}
&\psi_{n_A}(n)&=&\vert
\cala(p_1)\cdots\cala(p_{x-1})\;\cala(p_x)\cala(p_x+1)\;
\cala(p_{x+2})\cdots\cala(p_{n_A})\rangle\\
\longrightarrow&\psi_{n_A}(m)&=&\vert
\cala(p_1)\cdots\cala(p_{x-1})
\;\calb(p_x)\calb(p_x+1)\calb(p_x+2)\calb(p_x+3)\;
\cala(p_{x+2}+2)\cdots\cala(p_{n_A}+2)\rangle\\
&m&=&\ds\sum_{j=1}^{x-1}\binomial{p_j-1}{j}+\sum_{j=x+2}^{n_A}\binomial{p_j+1}{j-2}
\ea\ee
The same for a primed state $\psi_{n_A}'(n')$ with $n_A$ replaced by
$n_A-1$ in the last sum.

$p_x=L-n_A-2$ for a primed state:
%\nopagebreak
\be\ba{crcl}
&\psi_{n_A}'(n')&=&\vert
\cala(p_1)\cdots\cala(p_{n_A-2})\;\cala\cala\rangle'\\
\longrightarrow&\psi_{n_A}(m)&=&\vert
\calb\;\cala(p_1+1)\cdots\cala(p_{n_A-2}+1)\;\calb\calb\calb\rangle\\
&m&=&\ds\sum_{j=1}^{n_A-2}\binomial{p_j}{j}
\ea\ee

$p_x=L-n_A-1$ for an unprimed state:
%\nopagebreak
\be\ba{crcl}
&\psi_{n_A}(n)&=&\vert
\cala\;\cala(p_2)\cdots\cala(p_{n_A-1})\;\cala\rangle\\
\longrightarrow&\psi_{n_A}(m)&=&
\vert\calb\calb\;
\cala(p_2+1)\cdots\cala(p_{n_A-1}+1)\;\calb\calb\rangle\\
&m&=&\ds\sum_{j=2}^{n_A-1}\binomial{p_j}{j-1}
\ea\ee

$p_x=L-n_A$ for a primed state:
%\nopagebreak
\be\ba{crcl}
&\psi_{n_A}'(n')&=&\vert
\cala\;\cala(p_2)\cdots\cala(p_{n_A-1})\cala\;\rangle'\\
\longrightarrow&\psi_{n_A}(m)&=&\vert
\calb\calb\calb\;
\cala(p_2+2)\cdots\cala(p_{n_A-1}+2)\;\calb\rangle\\
&m&=&\ds\sum_{j=2}^{n_A-1}\binomial{p_j+1}{j-1}
\ea\ee

In order to determine the non-zero matrix elements for
creation/annihilation it is thus necessary to scan the whole particle
configuration corresponding to $\psi(n)$ and to record the allowed
values for $x$, $y$ and $p_x$ and $p_y$. An alternative technique
would be to try to store the (very sparse) matrix for ${\cal
  H}_{\rm eff}^{(2)}$. However, as in most cases, not the computational speed
but the storage requirement is the limiting factor for the maximum
possible system size.

\end{appendix}

\begin{figure}
\caption{\label{fig0}
  The one particle excitation energy $\Delta E(Q)$ for the free model
  from (\protect{\ref{exc}}) versus the contracted wave vector
  $Q\cdot(1+k_F/\pi)/2$ for different values of $k_F=\pi q/(1-q)$, with
  $q$ the domain wall density (which can be evaluated as a function of
  $\kappa$ and $\Gamma/J$ via equation (\protect{\ref{qeq}})). $k_F=0$
  corresponds to the FM-phase, $k_F=\pi/2$ corresponds to $q=1/3$, and
  for $k_F$ between $\pi/2$ and $\pi$ ($q=1/2$, i.e.\ $<\!2\!>$-phase)
  the evolution is simply reversed.
}
\end{figure}

\begin{figure}
\caption{\label{fig1}
  Results of exact diagonalization of the original ANNNI model
  (\protect{\ref{ham}})-(\protect{\ref{mast_eq}}) for L=16 and various values of
  $\Gamma/J$: Expectation value of the qunatity $E_{\rm
    class}/J+(1-\kappa)L$ defined in the text.}
\end{figure}

\begin{figure}
\caption{\label{fig2}
  Results of exact diagonalization of the complete model
  (\protect{\ref{eff_1t}})+(\protect{\ref{eff_2t}}). Shown are the
  results for the expectation value of the particle number in the
  ground state and the first excited state together with the gap
  $\Delta E$. It is L=24 and $\Gamma/J=1/20$.
}
\end{figure}

\begin{figure}
\caption{\label{fig3}
  The same as in fig.\ \protect{\ref{fig2}} but with $\Gamma/J=1/5$.
}
\end{figure}

\begin{figure}
\caption{\label{fig4a}
  Comparision of the structure function $f_q(\kappa)$, eq.\ 
  (\protect{\ref{struc}}), for the free and the complete model. It is
    L=24 and $\Gamma/J=1/20$. Note the different scale on the $y$-axis
    in the plot for $q=7/12$.}
\end{figure}

\begin{figure}
\caption{\label{fig4b}
  The same as in fig.\ \protect{\ref{fig4a}} but with $\Gamma/J=1/5$.
}
\end{figure}

\begin{figure}
\caption{\label{fig5}
  Comparision of the correlation function $C(r)$, eq.\ 
  (\protect{\ref{corr}}), of the complete model with $G(r)$ given 
  in eq.\ (\protect{\ref{crfit}}). 
  It is L=32 and $\Gamma/J=1/20$.
}
\end{figure}

\begin{figure}
\caption{\label{fig6}
  The same as in fig.\ \protect{\ref{fig5}} but with $\Gamma/J=1/5$.
}
\end{figure}

\begin{figure}
\caption{\label{fig7}
 Structure function $f_q(\kappa)$, eq.\ 
 (\protect{\ref{struc}}), for the complete model in the case L=32 and
 $\Gamma/J=1/5$.
}
\end{figure}

\begin{figure}
\caption{\label{fig8}
  Fit of (\protect{\ref{fit}}) to the correlation function $C(r)$ for
  $\kappa=0.4085$, the maximum of the $f_{q=1/12}$ curve shown in
  fig.\ \protect{\ref{fig7}}. It is L=32 and $\Gamma/J=1/5$. The
  fit parameters are $a=0.21$, $\xi=3.6$.
}
\end{figure}


\begin{references}
\bibitem{footnote} For the infinite system it will be enough
        to check $k=1$ and $k=N/2-1$, respectively, but for a periodic system
        we must use even $k$'s.
\end{references}
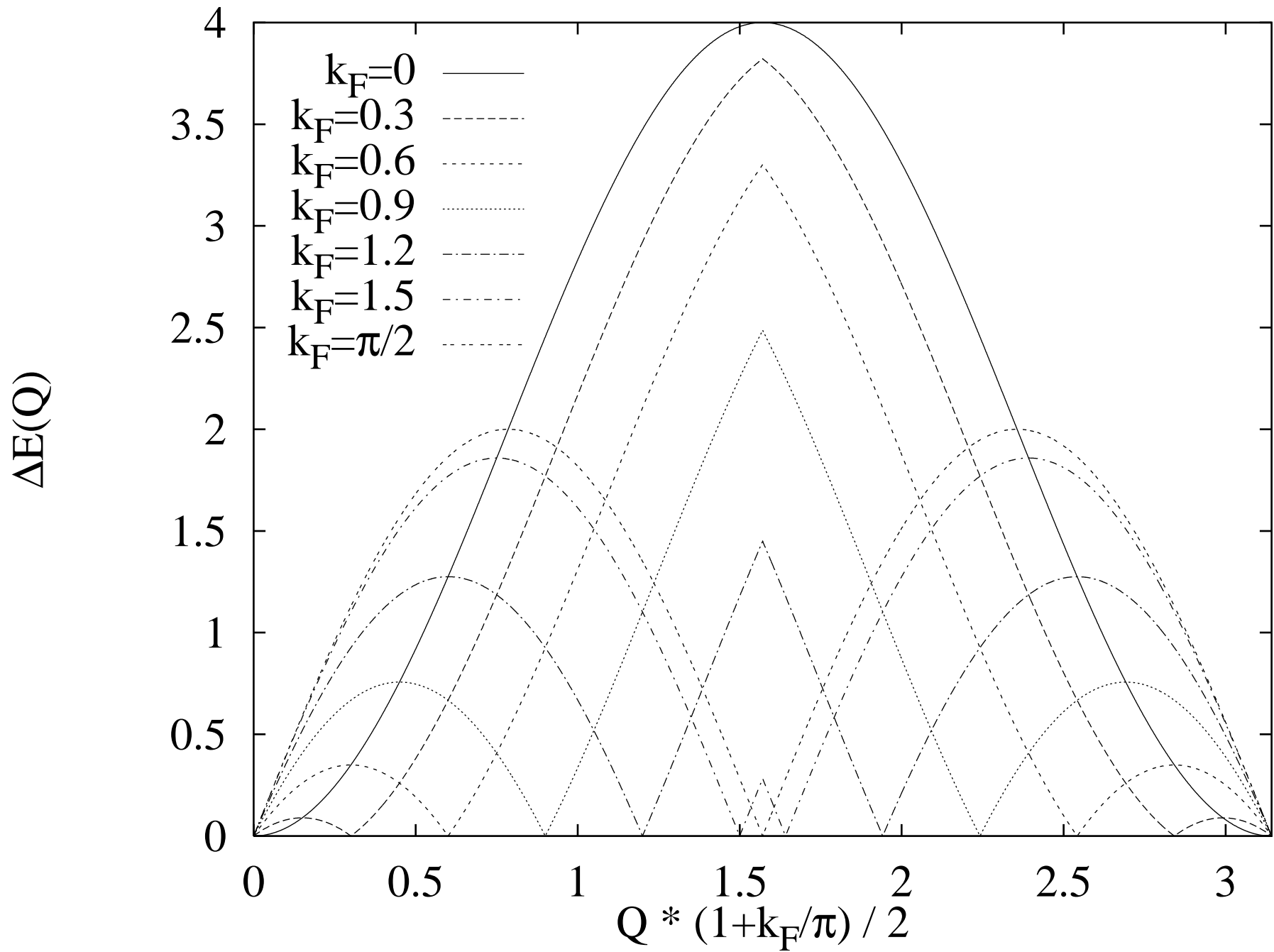
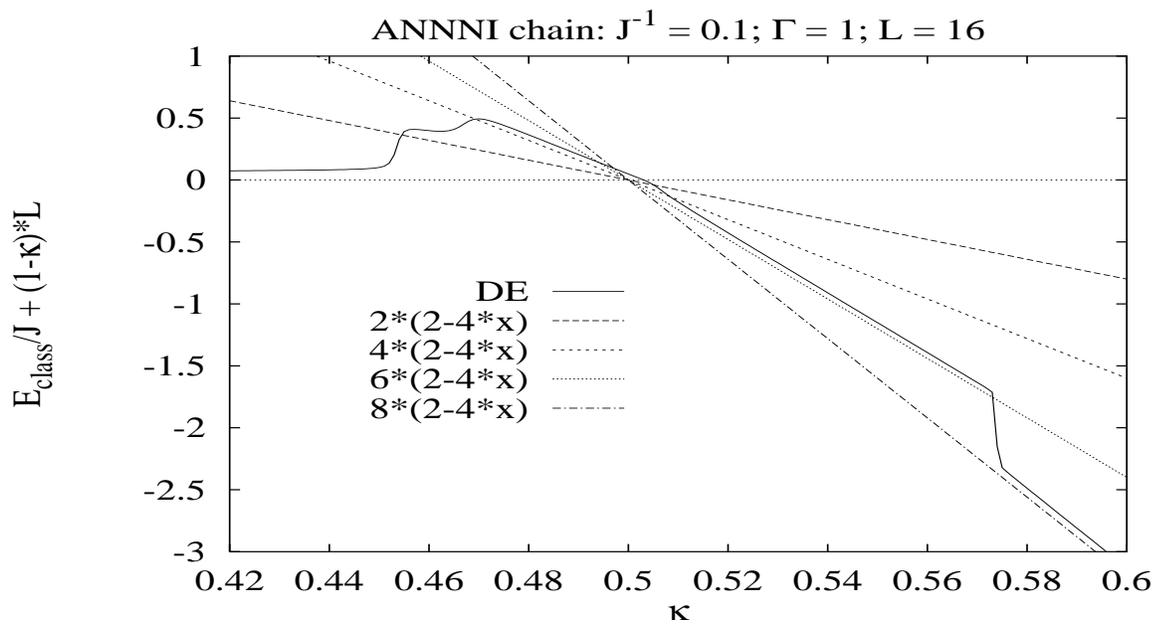
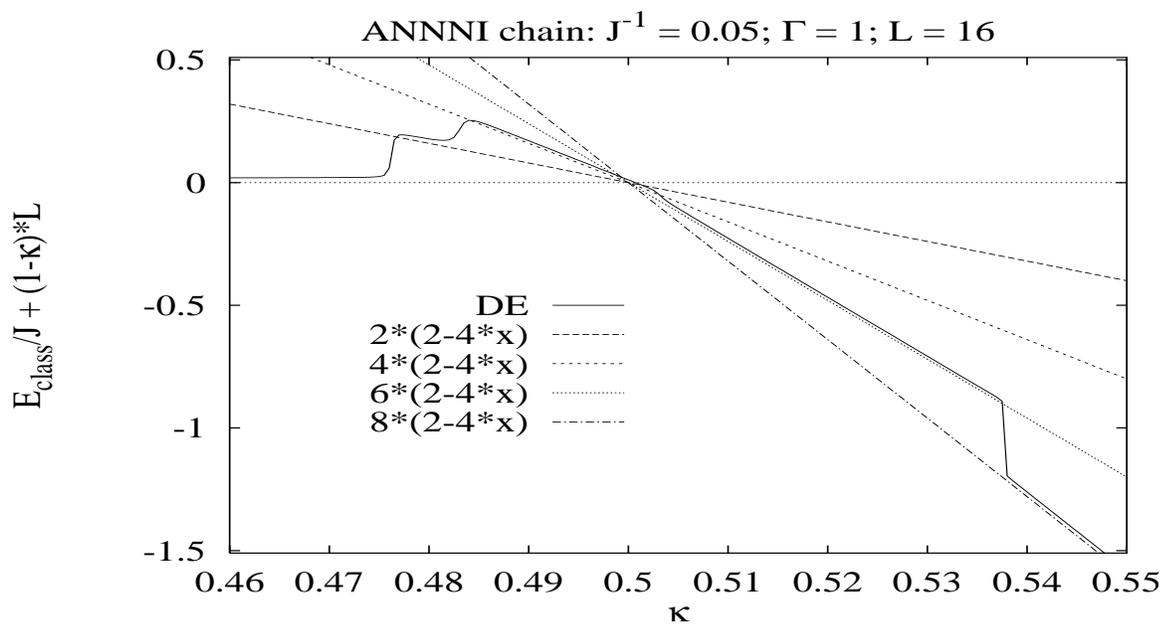
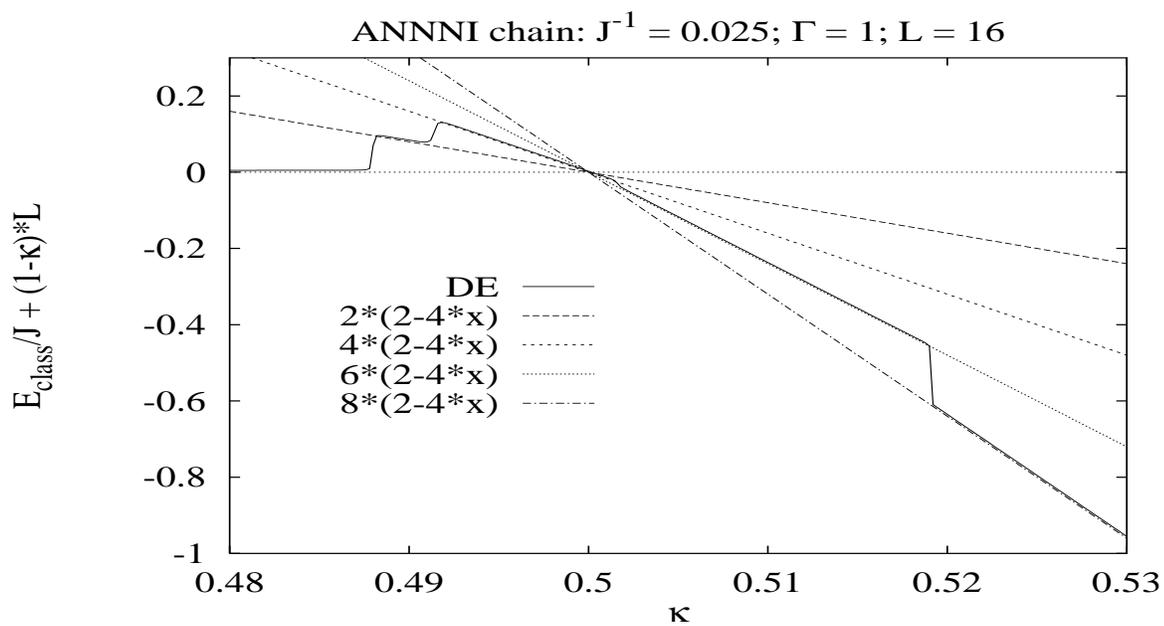
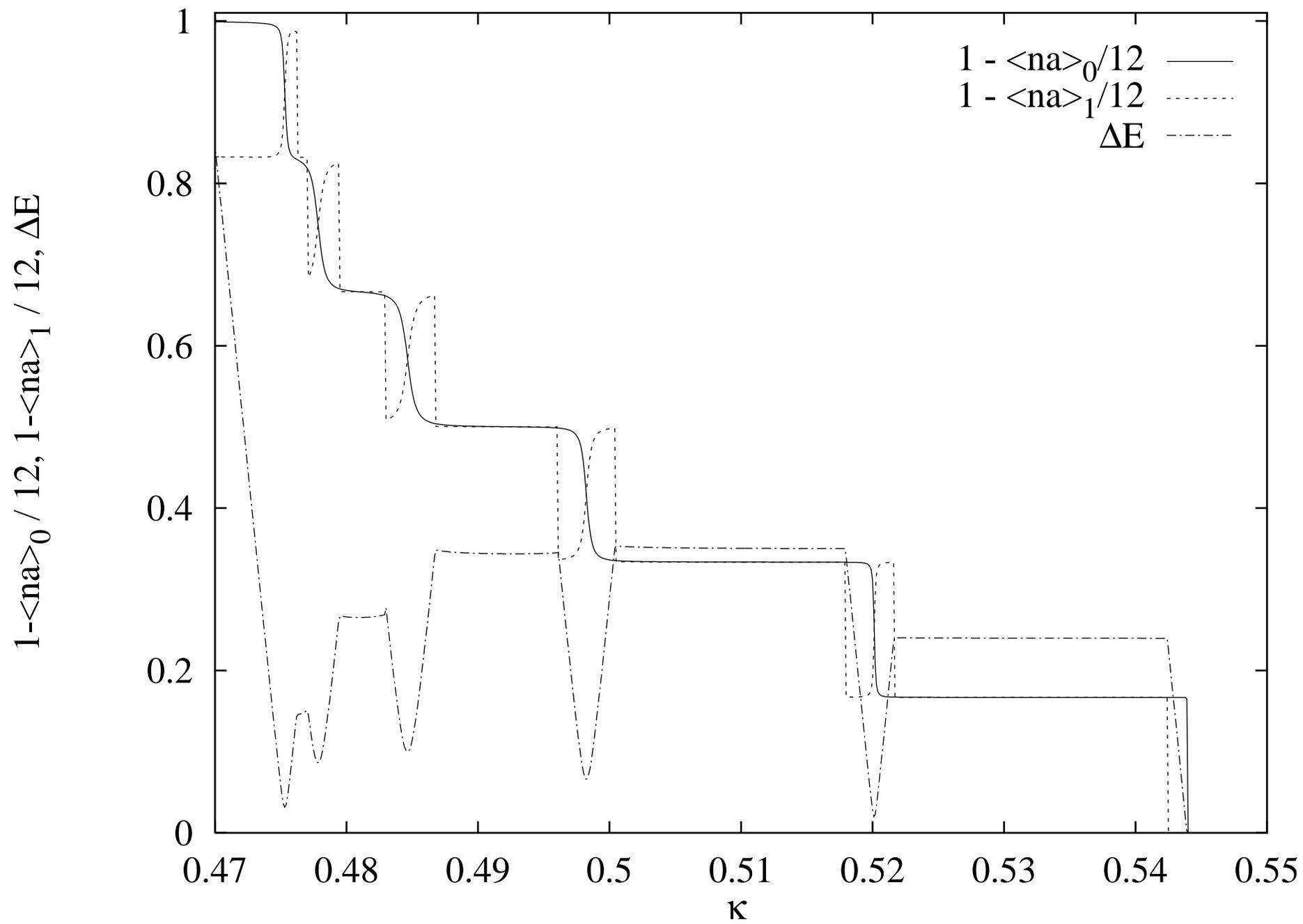
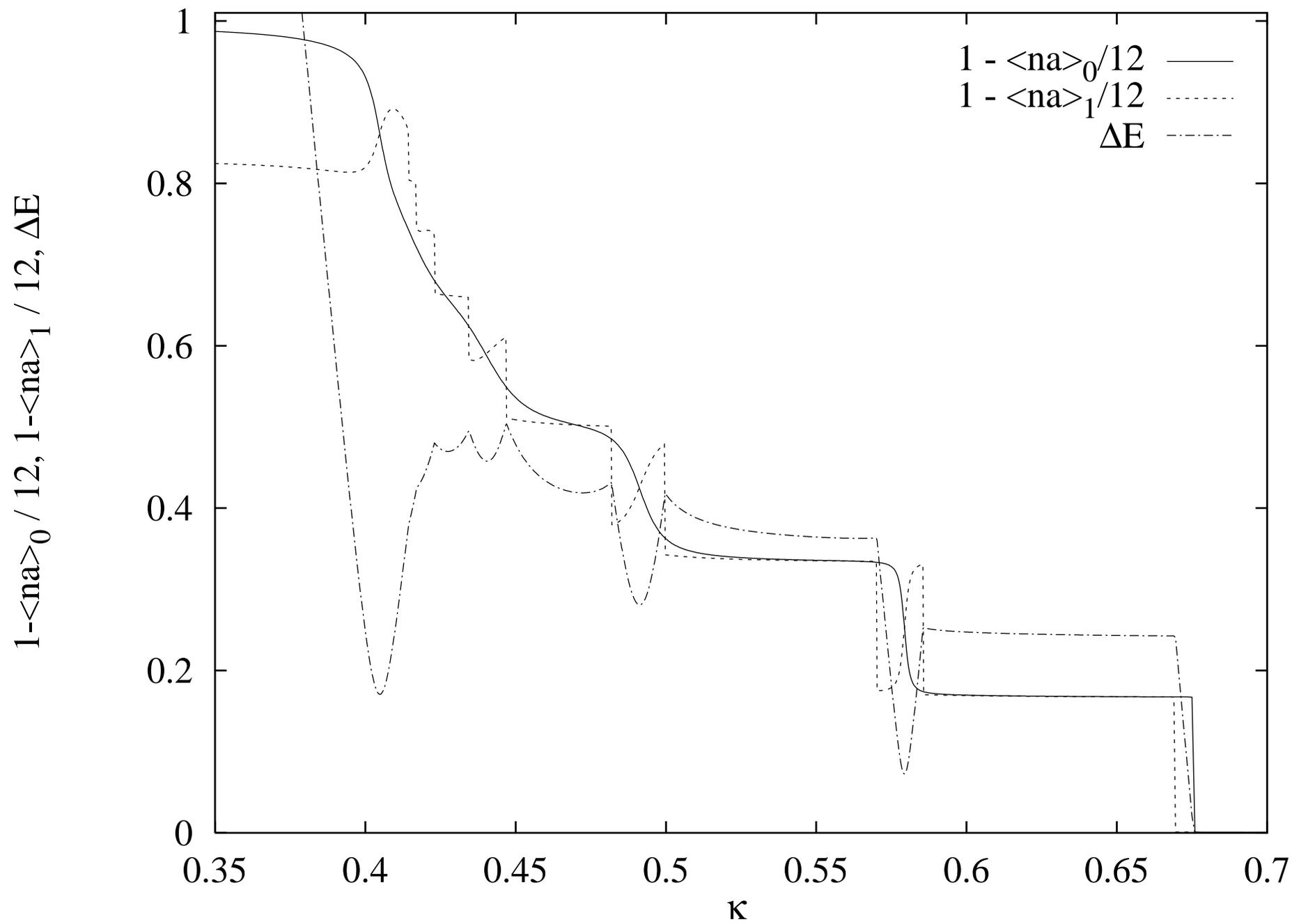
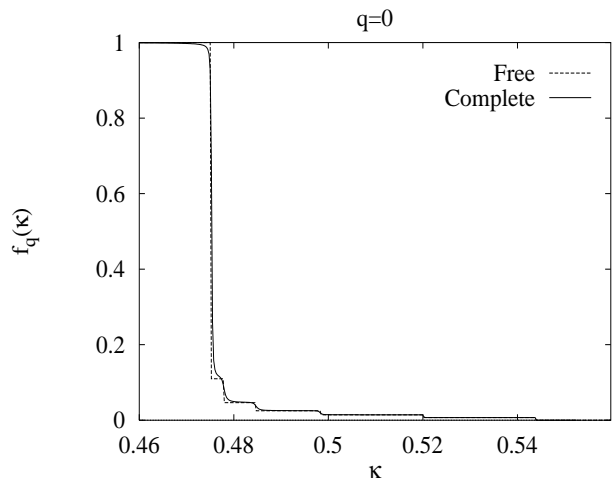
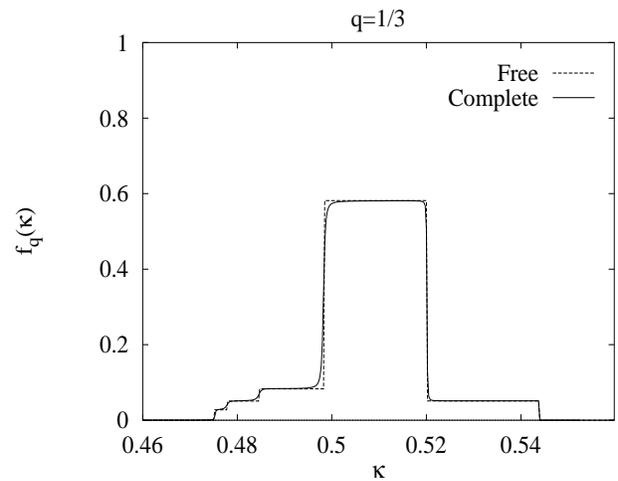
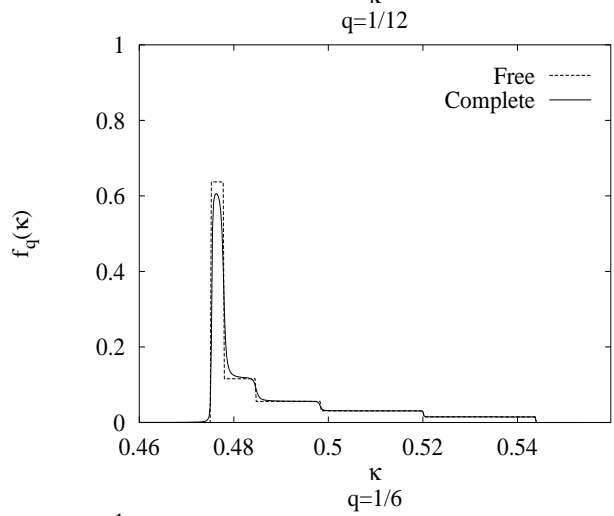
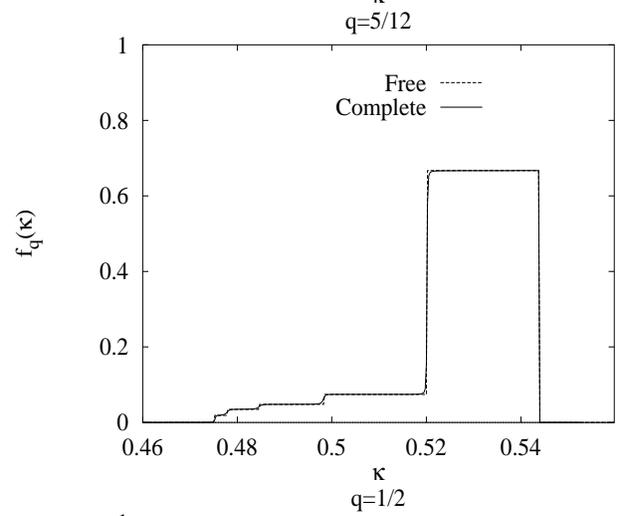
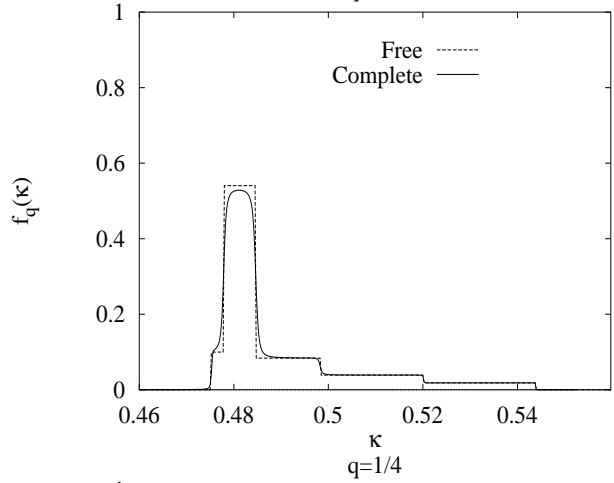
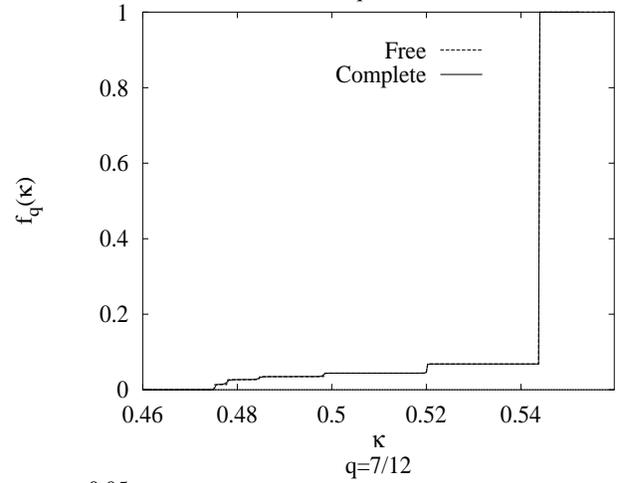
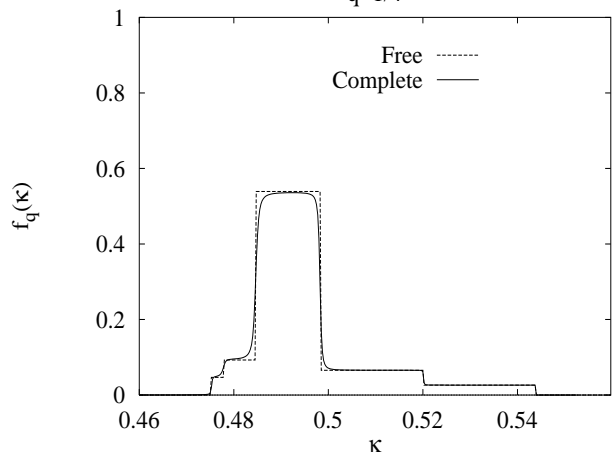
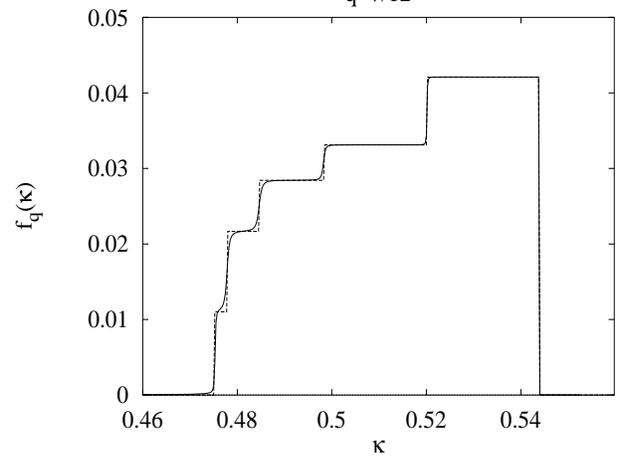
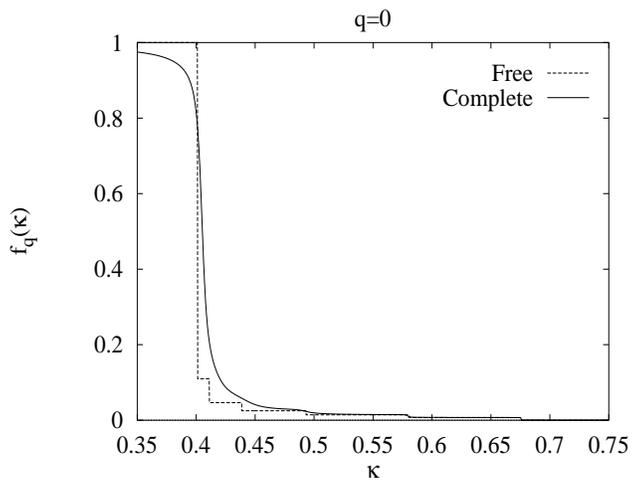
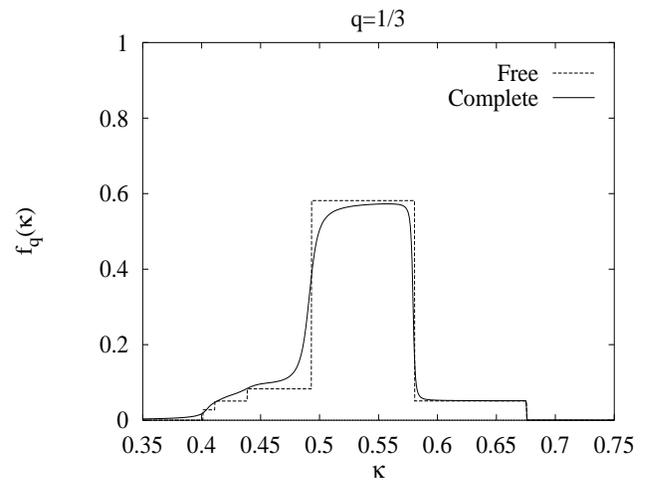
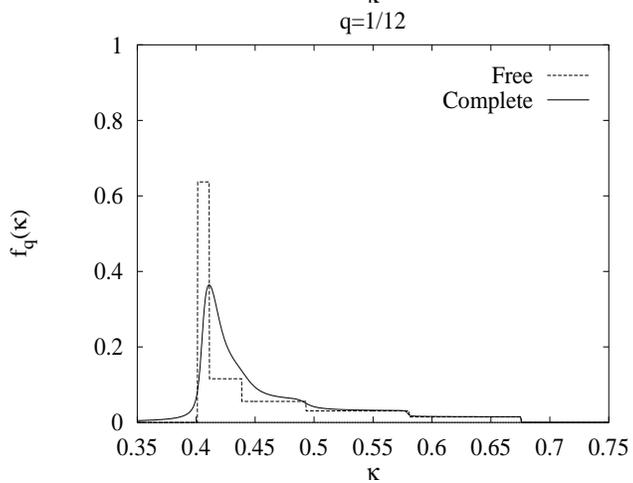
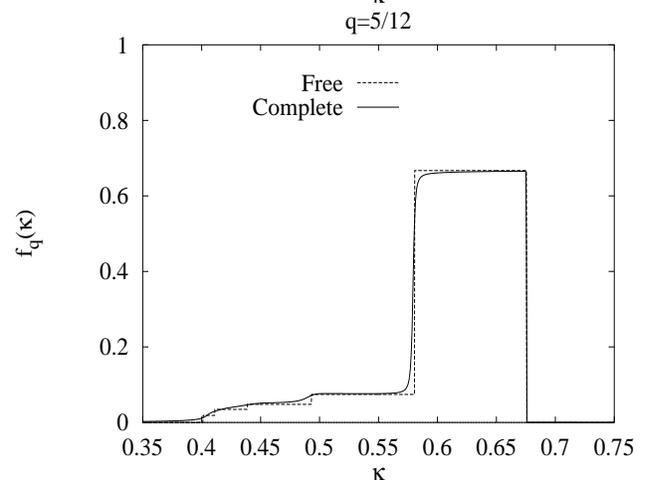
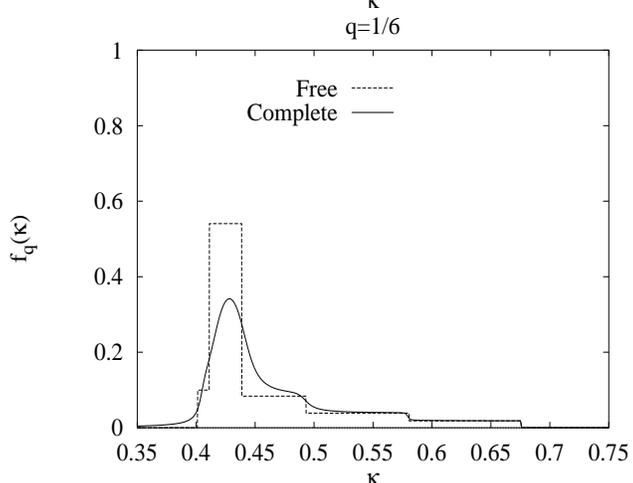
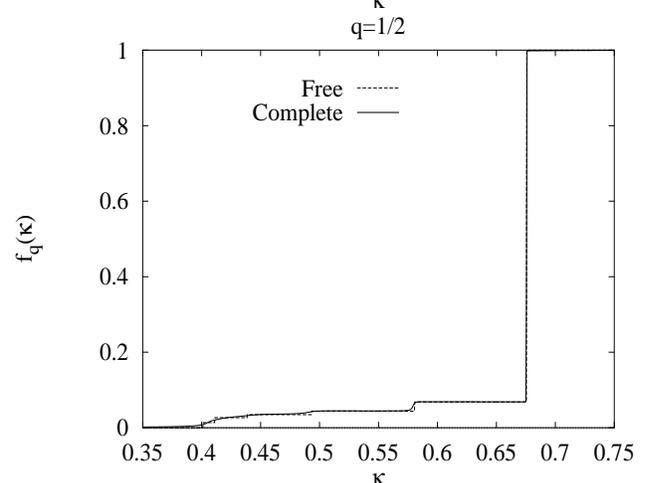
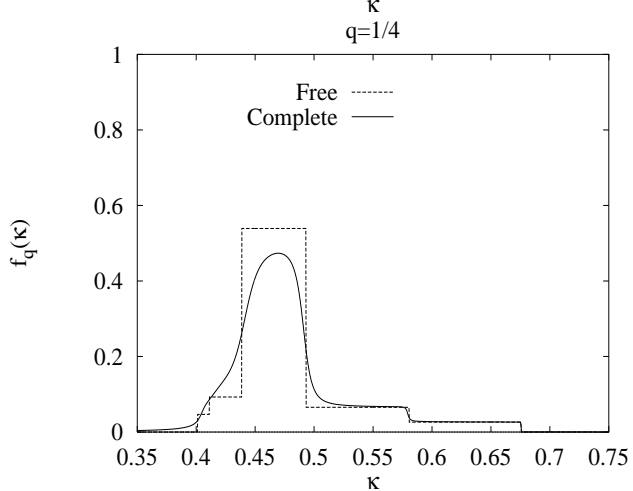
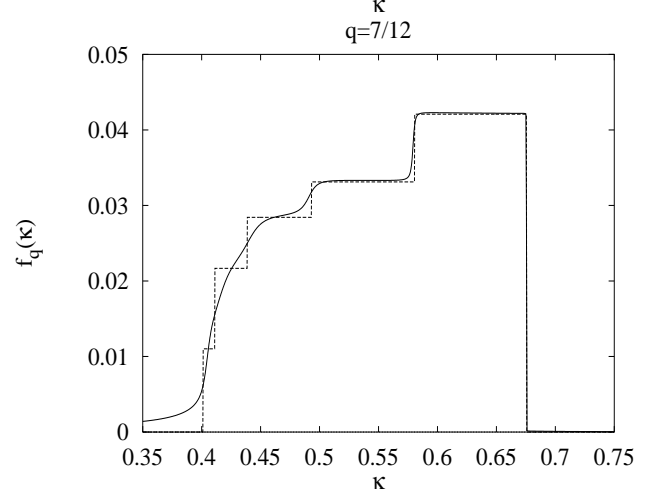
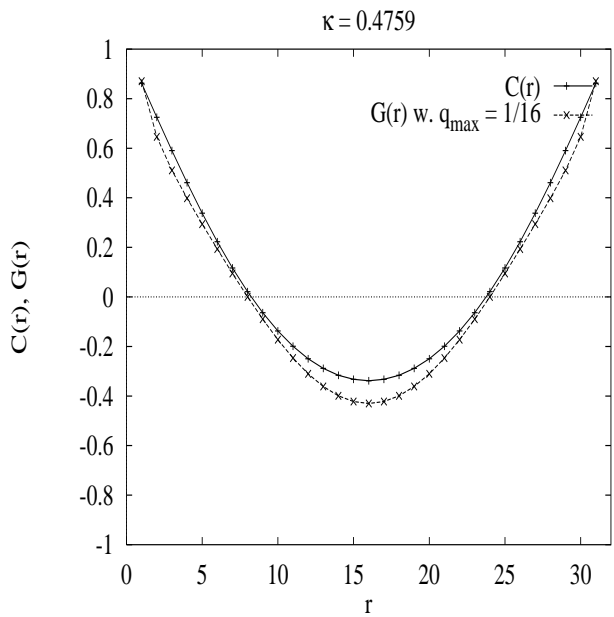
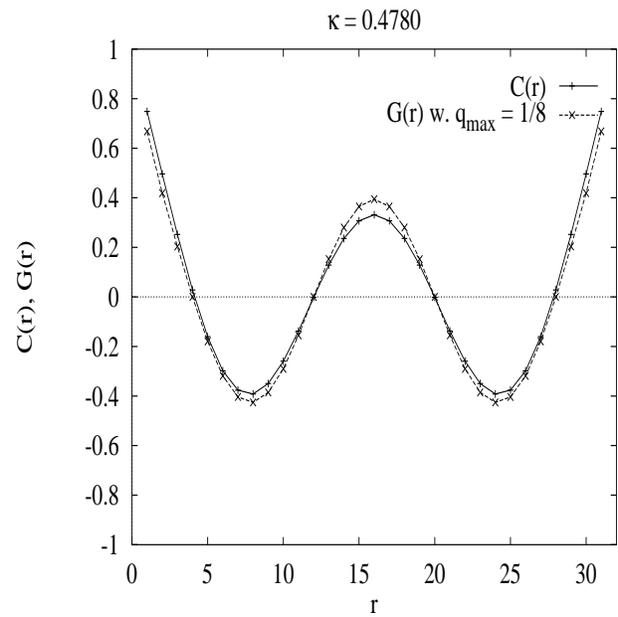
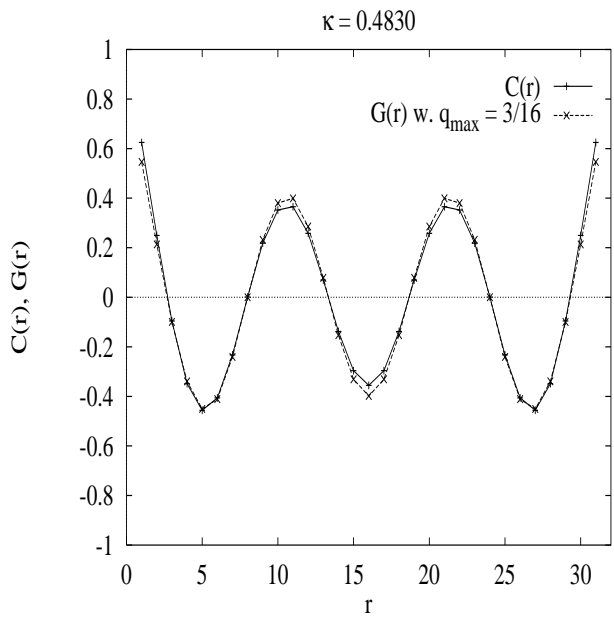
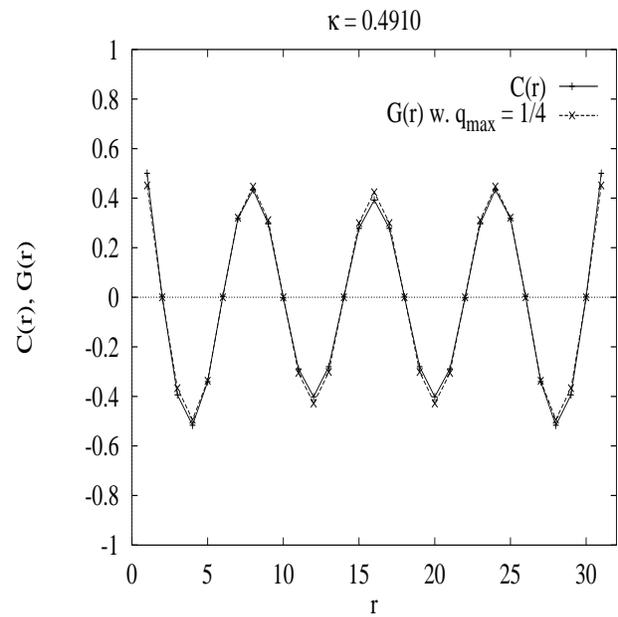
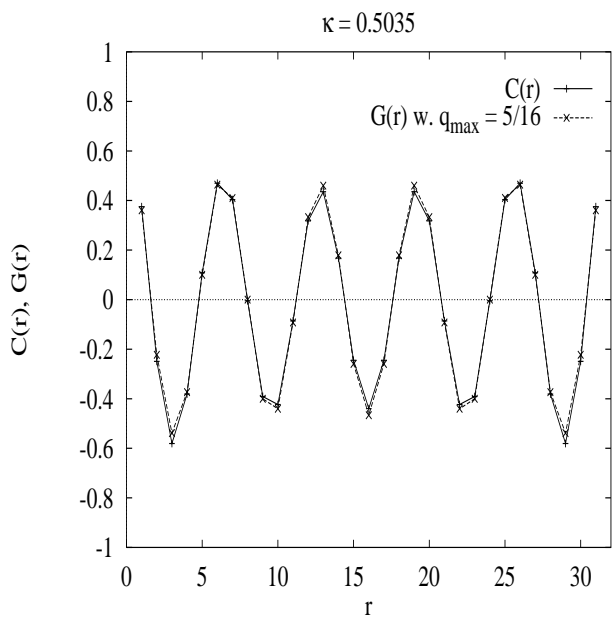
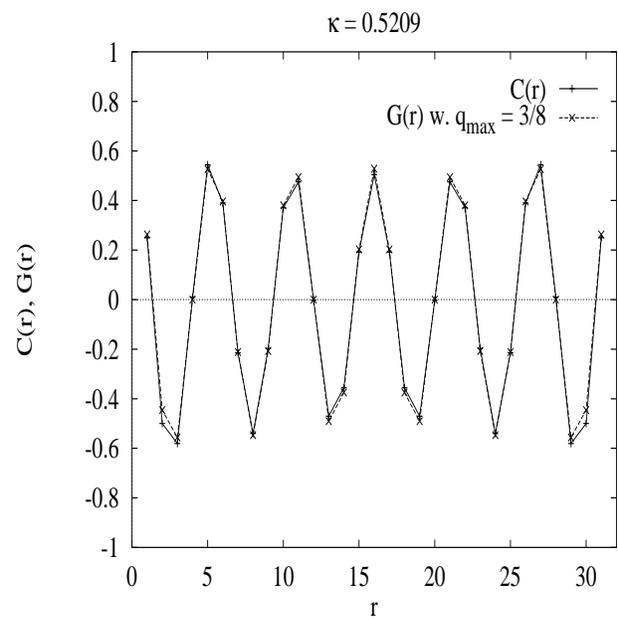
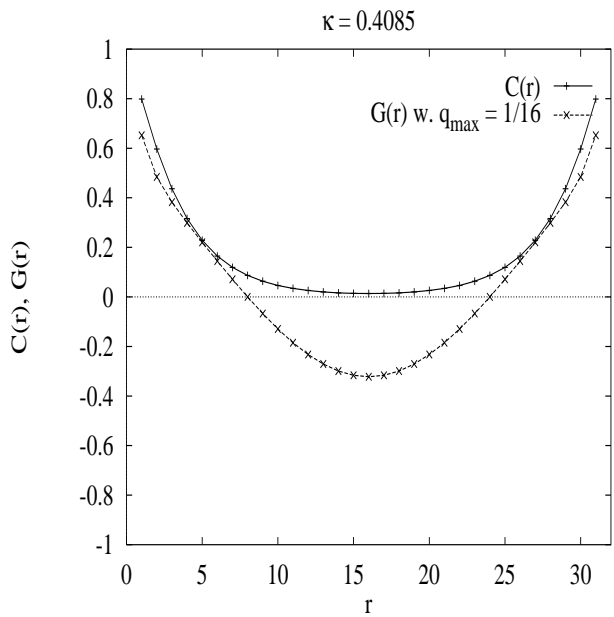
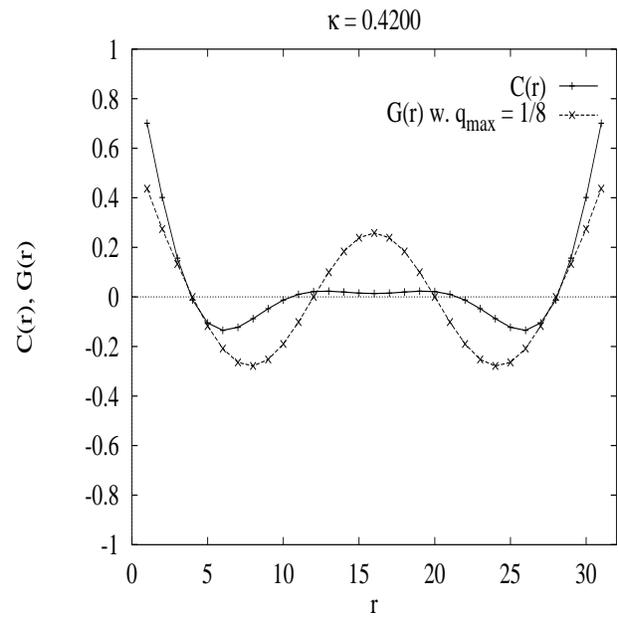
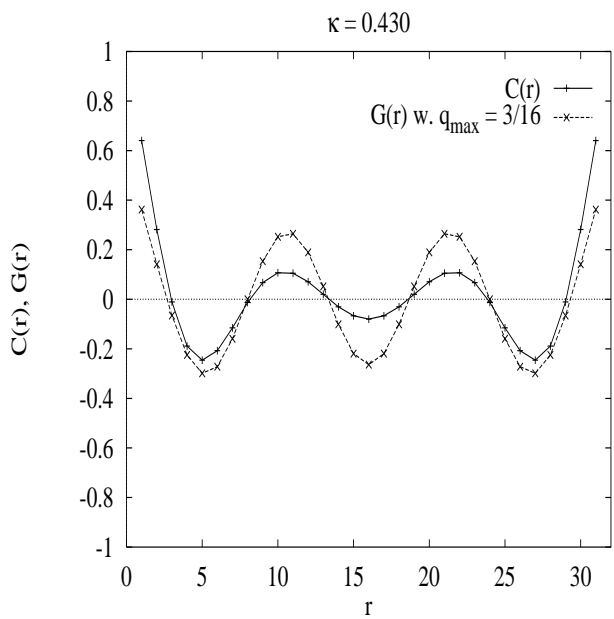
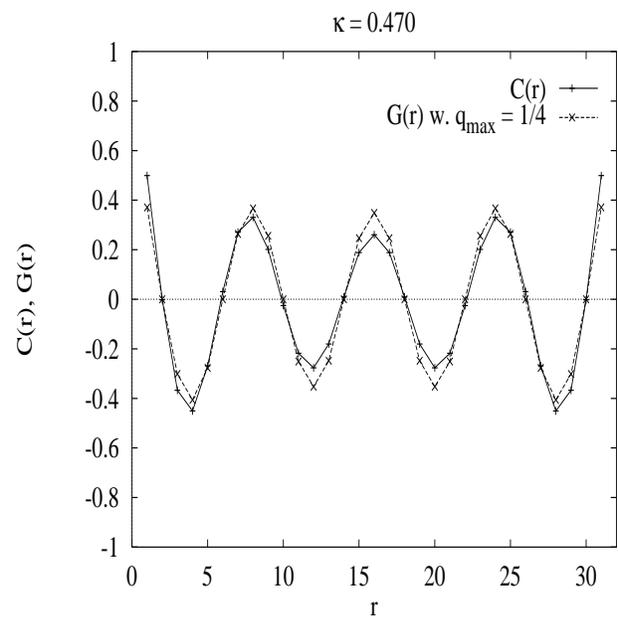
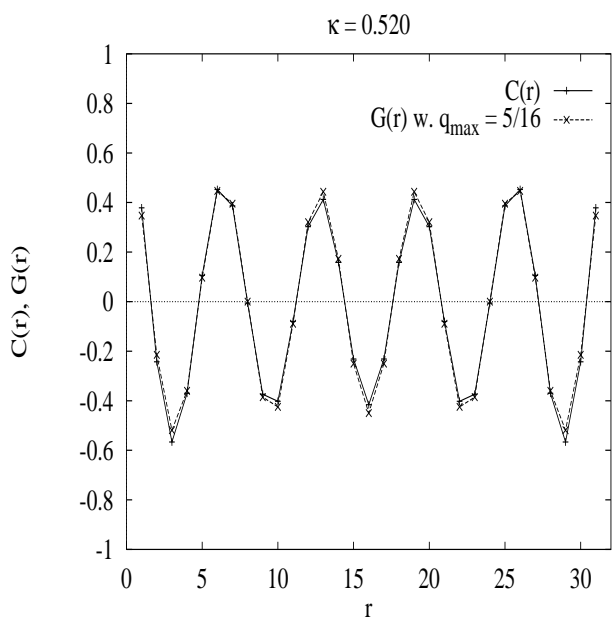
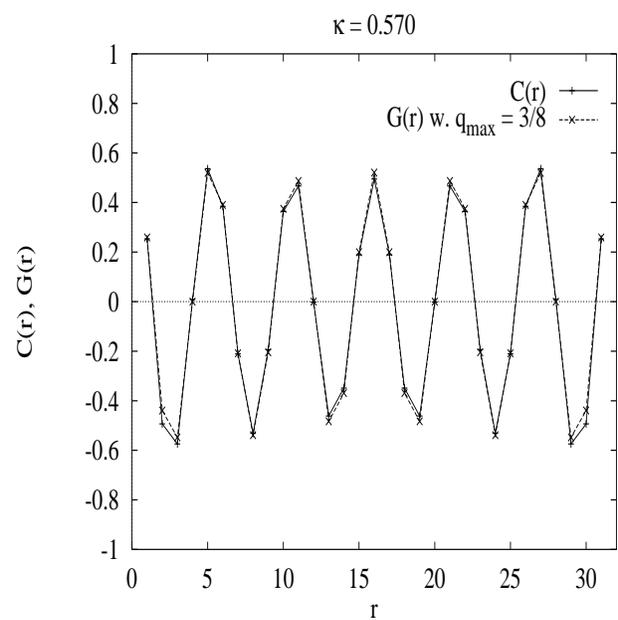
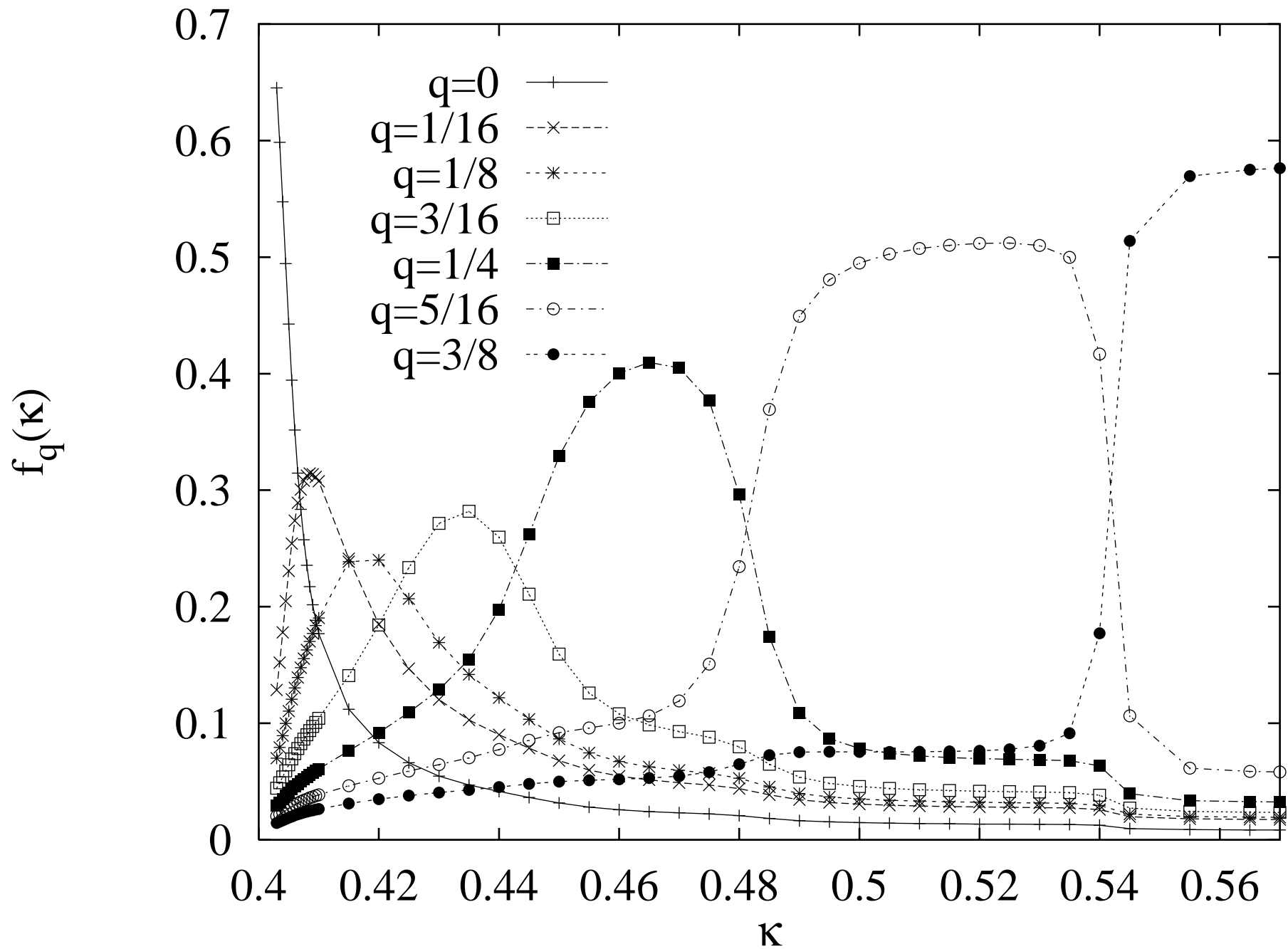
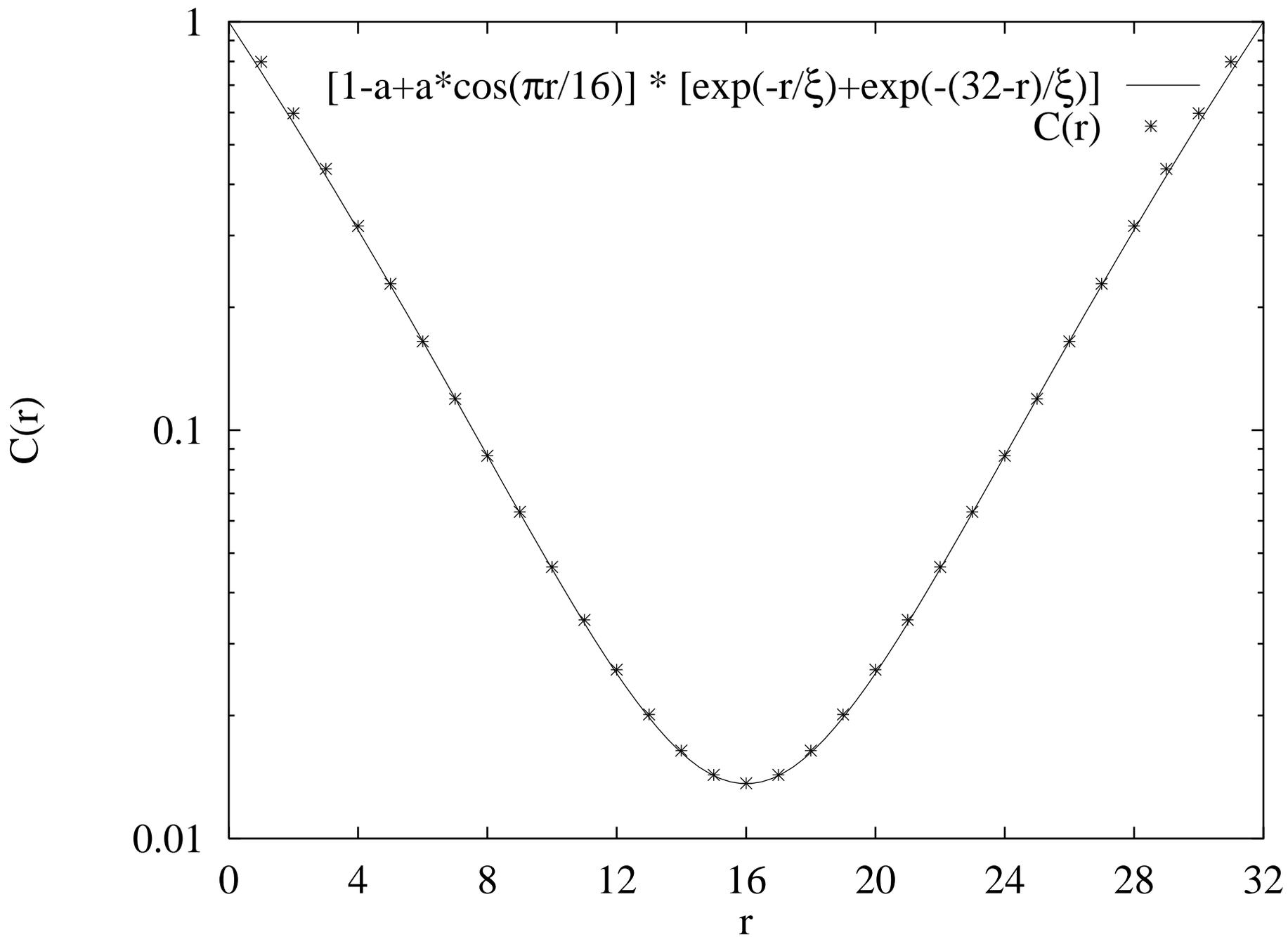
\end{document}